\documentclass[final,onefignum,onetabnum]{siamart190516}

\usepackage{mmap}

\usepackage{lipsum}
\usepackage{amsfonts}
\usepackage{graphicx}
\usepackage{epstopdf}
\usepackage[caption]{subfig} % was: subfignarms.sty
\usepackage{bm}
\usepackage{bbm}
\usepackage{amsmath}
\usepackage{multirow}
\usepackage{pdflscape}
\usepackage{mathtools}
\usepackage{tabularx}
\usepackage{overpic}
\usepackage{placeins}
\usepackage{caption}
\usepackage{amssymb}
\usepackage{cleveref}
\usepackage{lineno}
\usepackage{algorithm}
\usepackage[noend]{algpseudocode}
\usepackage{etoolbox}
\usepackage{tikz}
\usetikzlibrary{arrows.meta}
\tikzset{%
        >={Latex[width=2mm,length=2mm]},
        base/.style = {rectangle, rounded corners, draw=black,
                minimum width=2cm, minimum height=1.25cm,
                text centered,font=\rmfamily\footnotesize},
        output/.style = {base, fill=black!15},
        input/.style = {base, fill=green!15},
        truth/.style = {base, fill=black!35},
        process/.style = {base, fill=white}
}

\crefname{equation}{Eq.}{Eqs.}
\crefname{figure}{Fig.}{Figs.}
\crefname{tabular}{Tab.}{Tabs.}
\crefname{section}{Section}{Sections}
\crefname{subsection}{Subsection}{Subsections}
\crefname{algorithm}{Algorithm}{Algorithms}

\makeatletter
\newcommand*{\algrule}[1][\algorithmicindent]{%
        \makebox[#1][l]{%
                \hspace*{.2em}% <------------- This is where the rule starts from
                \vrule height .75\baselineskip depth .25\baselineskip
        }
}

\newcount\ALG@printindent@tempcnta
\def\ALG@printindent{%
        \ifnum \theALG@nested>0% is there anything to print
        \ifx\ALG@text\ALG@x@notext% is this an end group without any text?
        \else
        \unskip
        \ALG@printindent@tempcnta=1
        \loop
        \algrule[\csname ALG@ind@\the\ALG@printindent@tempcnta\endcsname]%
        \advance \ALG@printindent@tempcnta 1
        \ifnum \ALG@printindent@tempcnta<\numexpr\theALG@nested+1\relax
        \repeat
        \fi
        \fi
}
\patchcmd{\ALG@doentity}{\noindent\hskip\ALG@tlm}{\ALG@printindent}{}{\errmessage{failed to patch}}
\patchcmd{\ALG@doentity}{\item[]\nointerlineskip}{}{}{} % no spurious vertical space
\makeatother

\makeatletter
\def\algbackskip{\hskip-\ALG@thistlm}
\makeatother

\ifpdf
  \DeclareGraphicsExtensions{.eps,.pdf,.png,.jpg}
\else
  \DeclareGraphicsExtensions{.eps}
\fi

\newsiamremark{remark}{Remark}
\newsiamremark{hypothesis}{Hypothesis}
\crefname{hypothesis}{Hypothesis}{Hypotheses}
\newsiamthm{claim}{Claim}

\DeclareMathOperator*{\argmin}{arg\,min}

\usepackage{amsopn}

\makeatletter
\long\def\@makecaption#1#2{%
    \footnotesize
    \setlength{\parindent}{1.5pc}
  \ifx\@captype\@figtxt
    \vskip\abovecaptionskip
    \setbox\@tempboxa\hbox{{\normalfont\scshape #1}. {\normalfont\itshape #2}}%
    \ifdim \wd\@tempboxa >\hsize
      {\normalfont\scshape #1}. {\normalfont\itshape #2}\par
    \else
      \global\@minipagefalse
      \hb@xt@\hsize{\hfil\box\@tempboxa\hfil}%
    \fi
  \else
    \hbox to\hsize{\hfil{\normalfont\scshape #1}\hfil}%
    \setbox\@tempboxa\hbox{{\normalfont\itshape #2}}%
    \ifdim \wd\@tempboxa >\hsize
      {\normalfont\itshape #2}\par
    \else
     \global\@minipagefalse
      \hb@xt@\hsize{\hfil\box\@tempboxa\hfil}%
    \fi
    \vskip\belowcaptionskip
  \fi}
\makeatother

\renewcommand{\thepage}{B\arabic{page}}

\title{Sequential Active Learning of Low-Dimensional Model Representations for Reliability Analysis\thanks{Submitted to the journal's Computational Methods in Science and Engineering section May 3, 2021; accepted for publication (in revised form) January 19, 2022; published electronically May 24, 2022.
\URL{10.1137/21M1416758}
\funding{This work was supported by the German Research Foundation (DFG) through grant STR 1140/6-1 under SPP 1886.}}}

\author{Max Ehre\thanks{Engineering Risk Analysis Group, Technical University of Munich, Munich, Germany (\href{mailto:max.ehre@tum.de} {max.ehre@\allowbreak tum.de}, \href{mailto:iason.papaioannou@tum.de} {iason.\allowbreak papaioannou@tum.de}, \email{straub@tum.de}).}\orcid{https://orcid.org/0000-0001-5379-3364}
\and Iason Papaioannou\footnotemark[2]
\and Bruno Sudret\thanks{Chair of Risk, Safety and Uncertainty Quantification, ETH Z{\"u}rich, Z{\"u}rich, Switzerland\break (\email{sudret@ibk.eth.ch}).}
\and Daniel Straub\footnotemark[2]}

\headers{ACTIVE SEQUENTIAL SUBSPACE IMPORTANCE SAMPLING}{M. Ehre, I. Papaioannou, B. Sudret, and D. Straub}

\begin{document}

%siam_id=M141675
%CODEN=SJOCE3
\slugger{sisc}{2022}{44}{3}{B558--B584}
\maketitle

\setcounter{page}{558}

\begin{abstract}
To date, the analysis of high-dimensional, computationally expensive engineering models remains a difficult challenge in risk and reliability engineering. We use a combination of dimensionality reduction and surrogate modeling termed partial least squares--driven polynomial chaos expansion (PLS-PCE) to render such problems feasible. Standalone surrogate models typically perform poorly for reliability analysis. Therefore, in a previous work, we have used PLS-PCEs to reconstruct the intermediate densities of a sequential importance sampling approach to reliability analysis. Here, we extend this approach with an active learning procedure that allows for improved error control at each importance sampling level. To this end, we formulate an estimate of the combined estimation error for both the subspace identified in the dimension reduction step and the surrogate model constructed therein. With this, it is possible to adapt the training set so as to optimally learn the subspace representation and the surrogate model constructed therein.
The approach is gradient-free and thus can be directly applied to black box--type models.
We demonstrate the performance of this approach with a series of low- (2 dimensions) to high- (869 dimensions) dimensional example problems featuring a number of well-known caveats for reliability methods besides high dimensions and expensive computational models: strongly nonlinear limit-state functions, multiple relevant failure regions, and small probabilities of failure.
\end{abstract}



\begin{DOI}
10.1137/21M1416758
\end{DOI}

%%%%%%%%%%%%%%%%%%%%%%%%%%%%%%%%%%%%%%%%%%%%%
\section{Introduction and previous work}
\label{sec:intro}
An important challenge in the design, analysis, and maintenance of engineering systems is the management of the associated uncertainties.
%Accounting for and quantifying uncertainties relevant to these tasks is therefore vital in tackling them.
It is common practice to analyze engineering systems by employing computational models that aim at representing the physical processes relevant to the system in consideration.
These computational models take the form of an input-output mapping. Therein, uncertainty is represented by equipping the model input with an appropriate probabilistic model.
Undesirable system responses are defined through a limit-state function (LSF). % which takes values $1$ if a response leads to failure and $0$ otherwise.
Reliability analysis is concerned with quantifying the probability of failure, which can be expressed as a $d$-fold integral of the input probability mass over the failure domain defined by nonpositive values of the LSF, where $d$ is the number of uncertain model inputs (see section~\ref{sec:reliability}).
%By definition, estimating a probability of failure should amount to estimating the probability of a rare event implying the probability of failure will assume small values.
In engineering, target failure probabilities are typically small; hence, reliability analysis requires the estimation of rare event probabilities.
Reliability analysis approaches can be categorized into approximation (e.g., the first- and second-order reliability methods FORM and SORM \cite{Rackwitz1978,Fiessler1979,DerKiureghian2005}) and simulation methods. If the LSF is only weakly nonlinear and the input dimension of the model is moderate, FORM and SORM perform well even for small failure probabilities. The simplest simulation method is the Monte Carlo method \cite{Owen2013}. The Monte Carlo method performs well independent of the problem input dimension, however its performance deteriorates as the failure probability decreases if the computational budget is fixed.
Various techniques such as importance sampling (IS) \cite{Bucher1988,Engelund1993,Au1999} and line-sampling \cite{Hohenbichler1988, Koutsourelakis2004} have been proposed to mitigate this dependence on the magnitude of the failure probability.
More recently, sequential Monte Carlo methods such as subset simulation \cite{Au2001} and IS-based sequential methods \cite{Kroese2013, Kurtz2013, WangSong2016, Papaioannou2016, Rubinstein2017,  Papaioannou2019b} have been used successfully to efficiently solve high-dimensional reliability problems with small failure probabilties.
If the computational model is expensive and a hierarchy of increasingly coarse and cheap models is accessible, multilevel and multifidelity \cite{Peherstorfer2018} Monte Carlo methods can help alleviate computational cost by performing most model evaluations on the cheaper models (e.g., a discretized differential equation with coarser resolution). In \cite{Ullmann2015}, multilevel Monte Carlo is combined with subset simulation, and recently \cite{Wagner2020} introduced multilevel sequential IS based on the sequential IS approach in \cite{Papaioannou2016}.
All of the above-mentioned approaches are designed to work with the probabilistic computational model directly. However, often this model encompasses a numerical solver for (sets of) partial differential equations such that a model evaluation is computationally expensive.
%\enlargethispage*{1pc}

\looseness-1{}This has increasingly lead researchers to turn towards surrogate model--based reliability methods. Such methods attempt to approximate the expensive computational model with a cheap surrogate model, whose coefficients are identified based on a set of original model evaluations: the training set.
\cite{Faravelli1989} used a polynomial response surface method for performing reliability analysis as early as 1989. \cite{Guan2001} proposed an improved version of the response surface method. Since then, a variety of surrogate modeling techniques has been applied in the context of reliability analysis such as artificial neural networks  \cite{Papadrakakis1996,Hurtado2001,Schueremans2005}, support vector machines \cite{Hurtado2007,Bourinet2011,Bourinet2016}, Gaussian process regression-based models \cite{Echard2011,Dubourg2013}, and projection to polynomial bases including polynomial chaos expansions PCEs \cite{Li1998,Li2010,Li2011,Sudret2013} and low-rank tensor approximations \cite{Konakli2016c}.

\textit{Static}, \textit{global} surrogate models suffer from a decrease in accuracy in the tails of the model response distribution such that they are of limited use for reliability analysis.
%Thus, a large body of literature discussing possible remedies has formed.
In this context, \textit{static} refers to surrogate models that are constructed based on a fixed training set, and
\textit{global} refers to surrogate models that are trained and evaluated on the entire input space (as opposed to locally con- and re-fined models).
Thus, one can distinguish two strategies to overcome this limitation:
\begin{itemize}
        \item \textit{Locality}: Surrogate models are coupled with sequential sampling techniques which serve to focus the training set and accuracy in the relevant regions around the failure hypersurface \cite{Papadopoulos2012,Bourinet2011,Bourinet2016,Bect2017,Papaioannou2018b}.
        \item \textit{Adaptivity} (in the training set): The training set is augmented with points that are most informative with respect to the failure probability estimate according to an ``in-fill criterion.'' The refined surrogate model is then used to estimate the probability of failure with a sampling method and a large number of cheap samples. Such procedures are summarized under the term active learning (AL) or optimal experimental design. AL in combination with crude Monte Carlo have been applied in reliability-based optimization and reliability analysis in \cite{Echard2011,Oakley2004,Bichon2008,Picheny2010}.
        \cite{Schueremans2005} investigates the performance of splines and neural networks in combination with directional sampling and IS, and \cite{Dubourg2013,Cadini2014} combine Gaussian process models with IS. \cite{Schoebi2017} proposes a crude Monte Carlo procedure relying on a Gaussian process surrogate model with PCE-based mean trend (PCE-Kriging) along with a novel termination criterion for the AL.
\end{itemize}
Often, both AL and sequential sampling techniques are combined using various combinations of in-fill criteria and sequential sampling techniques such as adaptive IS \cite{Balesdent2013} and subset simulation \cite{Bourinet2011,Huang2016,Bect2017, Bourinet2016}.
\cite{Marelli2018} turns away from surrogate models that naturally provide a measure of prediction uncertainty such as Gaussian processes or support vector machines and demonstrate how an AL algorithm can be realized with PCE using a bootstrap estimator of the PCE prediction uncertainty.

In spite of a plethora of existing approaches to surrogate-assisted reliability analysis, the literature on high-dimensional problems ($d \ge 100$) in this context is scarce.
\cite{Jiang2017,Li2020} propose to perform reliability analysis with a static, global Kriging model constructed in a low-dimensional linear subspace of the original model input space, which is identified by the active subspaces method \cite{Constantine2014} and autoencoders, respectively. Both \cite{Jiang2017,Li2020} apply their methods to moderate-dimensional problems with up to $d=20$ and $d=40$ input variables, respectively.
\cite{Pan2017} uses sliced inverse regression to identify a linear low-dimensional subspace and construct a static, global PCE in this space based on which they perform reliability analysis directly.
\cite{Zhou2020} develops these ideas further by combining the active subspace-Kriging model with an AL approach and applies this combination to a high-dimensional analytical problem of $d=300$ that possesses a perfectly linear low-dimensional structure.

In this work, we propose an importance sampler based on a dimensionality-reducing surrogate model termed partial least squares--driven PCE (PLS-PCE) \cite{Papaioannou2019} to efficiently solve high-dimensional reliability problems with underlying computationally expensive, nonlinear models and small target probabilities ($\mathcal{O}(10^{-9})$).
Similar to sliced inverse regression and active subspaces, PLS-PCE achieves dimensionality reduction by identifying a low-dimensional linear subspace of the original input space. Our method is based on \cite{Papaioannou2018b} but introduces AL to refine the PLS-PCE approximation in each sequence of the IS procedure.
In \cite{Papaioannou2018b}, PLS-PCE models are reconstructed in each level of a sequential IS (SIS) scheme that is used to gradually shift the importance density towards the optimal importance density.
In this work, we augment this approach with two novel contributions to rare event simulation of computationally expensive, potentially (but not necessarily) high-dimensional and nonlinear models:
\begin{enumerate}
        \item We demonstrate how to perform AL with PCE models by deriving an in-fill criterion from large-sample properties of the PCE coefficient estimates.
        \item We use projection to linear subspaces to construct efficient surrogate models for high-dimensional problems and include the subspace estimation error in the in-fill criterion. This means, we are not only learning the surrogate model but also the subspace itself.
\end{enumerate}
Using AL in the context of PLS-PCE--based SIS  provides effective error control and benefits from the local confinement of the learning procedure of each subspace/\break{} surrogate model combination to the support of the current importance density.
%%%\added[id=2]
{Constructing local variance estimates for polynomial models in the way we propose here creates new possibilities to design goal-oriented surrogate modeling approaches that are driven by adaptive sampling based on such models (where so far, Gaussian processes were the dominant tool).}

In section~\ref{sec:reliability}, we set up the reliability problem and discuss the crude Monte Carlo sampler of the probability of failure. \cref{sec:sis} reviews IS and a variant of SIS \cite{Papaioannou2016} that is at the base of our approach. \cref{sec:plspce} introduces PLS-PCE models and their construction. \cref{sec:AL} details the theoretical foundations of AL of PLS-PCE models within SIS and summarizes our approach.
In section~\ref{sec:numerical_experiments}, we present comprehensive investigations of the method's performance in two engineering examples and provide a detailed discussion of the results. Conclusions are given in section~\ref{sec:conclusion}.
%%%%%%%%%%%%%%%%%%%%%%%%%%%%%%%%%%%%%%%%%%%%%
\section{Reliability analysis}
\label{sec:reliability}
Consider a system represented by the computational model $\mathcal{Y}: \mathbb{D}_{\bm{X}} \rightarrow \mathbb{R}$ with $d$-dimensional continuous random input vector $\bm{X}: \Omega \rightarrow \mathbb{D}_{\bm{X}} \subseteq \mathbb{R}^d$, where $\Omega$ is the sample space of $\bm{X}$ and by $F_{\bm{X}}(\bm{x})$, we denote its joint cumulative distribution function (CDF). $\mathcal{Y}$ maps to the system response $Y = \mathcal{Y}(\bm{x})$ with the model input $\bm{x} \in \mathbb{D}_{\bm{X}}$. Based on the response $Y$, unacceptable system states are defined by means of the LSF $\tilde{g}(Y)$. Defining $g(\bm{x}) = \tilde{g} \circ \mathcal{Y}(\bm{x})$ and introducing the convention
$$g(\bm{x}) = \begin{cases}
\le 0,\quad  \mathrm{failure}, \\
> 0,\quad  \mathrm{safety},
\end{cases}$$
the failure event of the system is defined as $\mathrm{F} = \{\bm{x} \in \mathbb{D}_{\bm{X}}:g(\bm{x}) \le 0\}$.
The probability of failure is given by \cite{Ditlevsen1996}
\begin{equation}
\label{e:pf}
p = \mathbb{P}(\mathrm{F}) = \int_{\mathbb{D}_{\bm{X}}} \mathrm{I}[g(\bm{x}) \le 0] f_{\bm{X}}(\bm{x})\mathrm{d}\bm{x} = \mathbb{E}_{f_{\bm{X}}}\left[\mathrm{I}(g(\bm{X}) \le 0)\right] ,
\end{equation}
where $ f_{\bm{X}}(\bm{x}) = \partial^d F/(\partial x_1\dots\partial x_d)|_{\bm{x}}$ is the joint probability density function of $\bm{X}$ and the indicator function $\mathrm{I}[\cdot]$ equals 1 if the condition in the argument is true and 0 otherwise.
Without loss of generality, one may formulate an equivalent reliability problem with respect to the standard-normal probability space using the random vector $\bm{U}: \Omega \rightarrow \mathbb{R}^d$. Given an isoprobabilistic transformation $T: \mathbb{D}_{\bm{X}} \rightarrow \mathbb{R}^d$ such that $\bm{U} = T(\bm{X})$ (see, e.g., \cite{Hohenbichler1981,Liu1986}), and defining $G(\bm{U}) = g(T^{-1}(\bm{U}))$, one can write \cref{e:pf} as
\begin{equation}
\label{e:pf2}
p = \int_{\mathbb{R}^d} \mathrm{I}[G\left(\bm{u}\right) \le 0] \varphi_d\left(\bm{u}\right)\mathrm{d}\bm{u} = \mathbb{E}_{\varphi_d}\left[\mathrm{I}(G(\bm{U}) \le 0)\right],
\end{equation}
where $\varphi_d$ denotes the $d$-dimensional independent standard-normal probability density function. The crude Monte Carlo estimate of \cref{e:pf2} is
\begin{equation}
\widehat{p}_{\mathrm{MC}} = \frac{1}{n} \sum\limits_{k=1}^{n} \mathrm{I}[G(\bm{u}^k) \le 0],~~~\bm{u}^k \stackrel{i.i.d.}{\sim}\varphi_d,
\end{equation}
where $\bm{u}^k \stackrel{i.i.d.}{\sim}\varphi_d$ means that $\{\bm{u}^k\}_{k=1}^n$ are $n$ samples that are independent and identically distributed according to $\varphi_d$.
This estimate is unbiased and has coefficient of variation (CoV)
\begin{equation}
\delta_{\mathrm{MC}}
= \sqrt{\frac{1-p}{np}}.
\end{equation}
The number of samples required to compute $\widehat{p}_{\mathrm{MC}}$ at a prescribed CoV $\delta_0$ reads
\begin{equation}
n_0 = \frac{1-p}{\delta_0^2 p} \stackrel{p \ll 1}{\approx} \frac{1}{\delta_0^2 p}.
\end{equation}
Therefore, crude Monte Carlo is inefficient for estimating rare event probabilities as, by definition, $p \ll 1$, and thus $n_0$ becomes large.

\section{SIS for rare event estimation}
\label{sec:sis}
Variance reduction techniques can be used to reduce the CoV of the probability estimate at a fixed budget of samples compared to crude Monte Carlo.
One of the most commonly used variance reduction methods is the IS method.
Let $h$ be a density such that $h\left(\bm{u}\right) > 0$ whenever $G\left(\bm{u}\right) \le 0$. Then, one can rewrite \cref{e:pf2}:
\begin{equation}
\label{e:pf_is}
p =\int_{\mathbb{R}^d} \mathrm{I}(G\left(\bm{u}\right) \le 0) \overbrace{\frac{\varphi_d\left(\bm{u}\right)}{h\left(\bm{u}\right)}}^{\omega \left(\bm{u}\right)} h\left(\bm{u}\right) \mathrm{d}\bm{u}
= \mathbb{E}_{h}\left[\mathrm{I}(G(\bm{U}) \le 0) \omega(\bm{U})\right],
\end{equation}
which leads to the (unbiased) IS estimator
\begin{equation}
\label{eq:phat_is}
\widehat{p}_{\mathrm{IS}} = \frac{1}{n} \sum\limits_{k=1}^{n} \mathrm{I}[G(\bm{u}^k) \le 0] \omega(\bm{u}^k),~~~\bm{u}^k \stackrel{i.i.d.}{\sim} h.
\end{equation}
The efficiency of IS depends intimately on the choice of the IS density $h$, and numerous techniques to construct it have been put forward. There exists an optimal importance density $h^*$ in the sense that it leads to $\mathbb{V}[\widehat{p}_{\mathrm{IS}}] = 0$:
\begin{equation}
\label{eq:hopt}
h^*\left(\bm{u}\right) = \frac{1}{p}\mathrm{I}[G\left(\bm{u}\right) \le 0] \varphi_d\left(\bm{u}\right).
\end{equation}
While this result is not immediately useful in estimating $p$ as it requires knowledge of $p$, it can be used to
guide the selection of a suitable IS function $h$.

The SIS method proposed in \cite{Papaioannou2016} selects the IS density sequentially starting from a known distribution $h_0$ that is easy to sample from. It relies on a sequence of distributions $\{h_i\left(\bm{u}\right)\}_{i=0}^M$,
\begin{equation}
\label{eq:eta_i}
h_i\left(\bm{u}\right) = \frac{\eta_i\left(\bm{u}\right)}{p_i},~i=1,\dots,M,
\end{equation}
where $\{\eta_i\left(\bm{u}\right)\}_{i=0}^M$ are nonnormalized versions of $\{h_i\left(\bm{u}\right)\}_{i=0}^M$ and $\{p_i\}_{i=0}^M$ are the respective normalizing constants.
The goal is to arrive at $h_M$, which is sufficiently close to $h^*$ based on some criterion, and perform IS with $h_M$. To this end, it is necessary to estimate $p_M$ and obtain samples from $h_M$.
Based on the likelihood ratio of two succeeding nonnormalized distributions $\omega_i\left(\bm{u}\right) = \eta_i\left(\bm{u}\right)/\eta_{i-1}\left(\bm{u}\right)$, we have
\begin{equation}
\label{eq:s_i}
s_i = \frac{p_i}{p_{i-1}} = \int_{\mathbb{R}^d}\frac{\eta_i\left(\bm{u}\right)}{\eta_{i-1}\left(\bm{u}\right)}h_{i-1}\left(\bm{u}\right) \mathrm{d}\bm{u} = \mathbb{E}_{h_{i-1}}\left[\omega_i\left(\bm{u}\right)\right].
\end{equation}
Therefore, an estimate of $p_M$ is given by
\begin{equation}
\label{eq:P_m_hat}
\widehat{p}_M = \prod_{i=1}^M \widehat{s}_i ~\mathrm{with}~ \widehat{s}_i = \frac{1}{n}\sum_{k=1}^{n} \omega_i\left(\bm{u}^k\right)
,~~~\bm{u}^k \stackrel{i.i.d.}{\sim} h_{i-1}.
\end{equation}
Samples from $h_i$ can be obtained using Markov chain Monte Carlo (MCMC) methods given samples from $h_{i-1}$. More precisely, \cite{Papaioannou2016} proposes a resample-move scheme in which Markov chain seeds are obtained as samples from $h_{i-1}$ that are then reweighted (resampled with weights) according to $\omega_i\left(\bm{u}\right)$. In this way, the seed samples are already approximately distributed according to the stationary distribution of the Markov chain $h_i$ and long burn-in periods can be avoided. We adopt an adaptive conditional MCMC sampler (aCS) to perform the move step due to its robust performance in high-dimensional settings. Details can be found in \cite{Papaioannou2016}.

\begin{figure}[t!]
\vspace*{-7pt}
        \centering
        \subfloat{\includegraphics[height=0.3\textwidth]{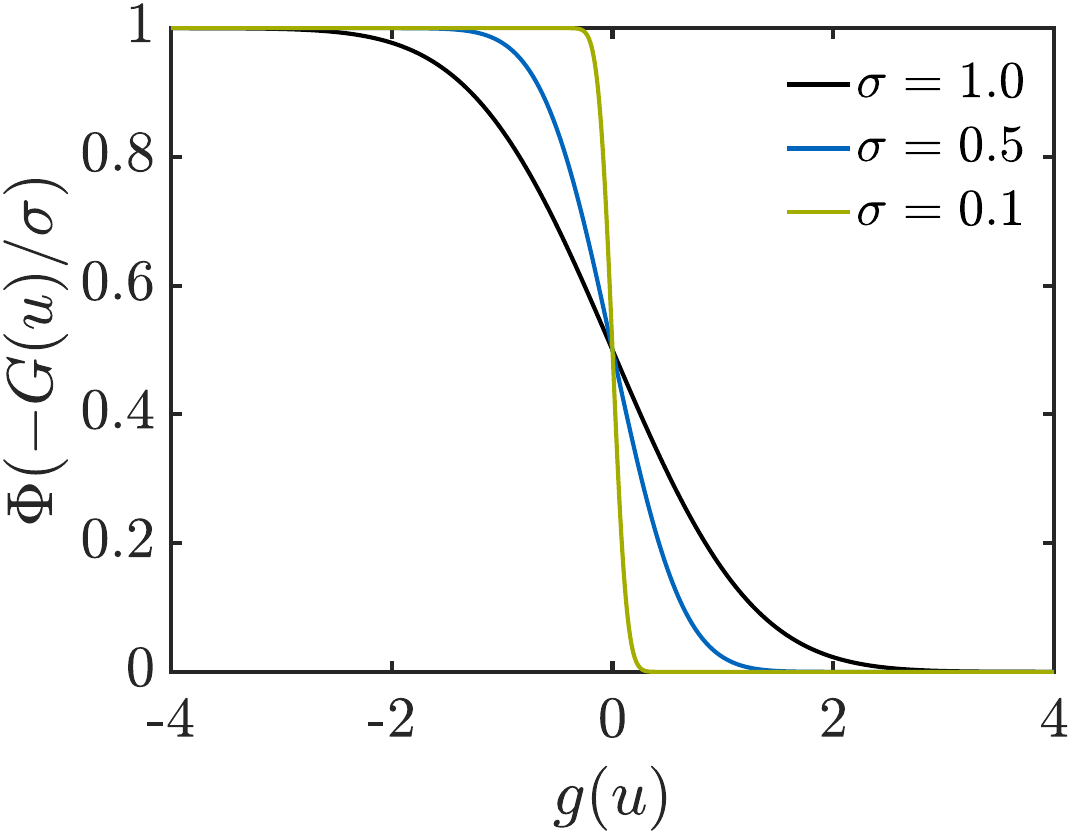}}\hspace{1.5cm}
        \subfloat{\raisebox{-5px}{\includegraphics[height=0.303\textwidth]{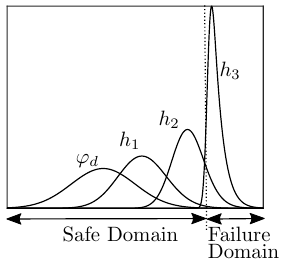}}}
        \caption{Smooth approximations to the indicator function $I(g(\bm{u})\le0)$ (left) and importance densities $h_i\left(\bm{u}\right) \propto \Phi\left(-G\left(\bm{u}\right)/\sigma_i \right)\varphi_d\left(\bm{u}\right)$ based on this approximation (right).}
        \label{fig:sis_schematic}
\end{figure}

The $h_i$ are chosen as smooth approximations of $h^*$ using the standard-normal CDF $\Phi(\cdot)$ (compare Figure~\ref{fig:sis_schematic}):
\begin{equation}
\label{eq:h_i}
h_i\left(\bm{u}\right) = \frac{1}{p_i} \Phi\left(-\frac{G\left(\bm{u}\right)}{\sigma_i} \right) \varphi_d\left(\bm{u}\right) = \frac{1}{p_i} \eta_i\left(\bm{u}\right),
\end{equation}
where $p_i = \mathbb{E}_{\varphi_d}[\Phi(-G(\bm{U})/\sigma_i)]$ is a normalizing constant and $\sigma_i$ is the smoothing parameter. Prescribing $\sigma_0 > \sigma_1 > \dots > \sigma_M$ ensures that the sequence $\{h_i\left(\bm{u}\right)\}_{i=0}^M$ approaches $h^*$.
% Figure 1
In each level, to avoid degeneration of the weights $\omega_i$ (meaning $\omega_i$ assuming values close to $0$ at all current samples), $h_{i-1}\left(\bm{u}\right)$ and $h_i\left(\bm{u}\right)$ cannot be too different in the sense that they share no support regions on which both have considerable probability mass. This is avoided by prescribing an upper bound for the estimated CoV of the weights $\widehat{\delta}_{w,i}= \widehat{\mathbb{COV}}[\omega_i(\bm{U})]$, which provides a criterion for determining $\sigma_i$:
\begin{equation}
\label{eq:sigma_i}
\sigma_i = \argmin\limits_{\sigma \in [0, \sigma_{i-1}]} \left(\widehat{\delta}_{\omega,i}(\sigma) - \delta_{\mathrm{target}}\right)^2.
\end{equation}
\cite{Papaioannou2016} recommends $\delta_{\mathrm{target}} = 1.5$. The algorithm terminates when $h_i$ is close enough to $h^*$ in the sense that
\begin{equation}
\label{eq:exit_crit}
\widehat{\mathbb{COV}}\left[\frac{h^*(\bm{U})}{h_i(\bm{U})}\right] =  \widehat{\mathbb{COV}}\left[\frac{\varphi_d(\bm{U})\mathrm{I}(G(\bm{U}) \le 0)}{\varphi_d(\bm{U}) \Phi(-G(\bm{U})/\sigma_i)}\right] =
\widehat{\mathbb{COV}}\left[\frac{\mathrm{I}(G(\bm{U}) \le 0)}{\Phi(-G\left(\bm{u}\right)/\sigma_i)}\right] \le \delta_{\mathrm{target}}.
\end{equation}
The final estimate of $\mathbb{P}(\mathrm{F})$ reads
\begin{equation}
\label{eq:pf_sis}
\widehat{p}_{\mathrm{SIS}}
= \widehat{p}_M \widehat{\mathbb{E}}_{\varphi_d}\left[\frac{\mathrm{I}(G(\bm{U}) \le 0)}{\eta_M(\bm{U})}\right]
= \left(\prod_{i=1}^{M}\widehat{s}_i\right) \frac{1}{n} \sum_{k=1}^{n}\frac{\mathrm{I}(G(\bm{u}^k) \le 0)}{\Phi(-G(\bm{u}^k)/\sigma_M)},~~~\bm{u}^k \stackrel{i.i.d.}{\sim} h_M.
\end{equation}
\cref{alg:SIS-aCS} summarizes the complete SIS-aCS procedure.
\begin{algorithm}
        \caption{SIS-aCS \cite{Papaioannou2016}} \label{alg:SIS-aCS}
        \begin{algorithmic}[1]
                \State \textbf{Input} {LSF $G\left(\bm{u}\right)$, target CoV $\delta_{\mathrm{target}}$, samples per level $n$, input dimension $d$, burn-in period $b$, max. \hspace*{1.05cm} iterations $i_{\mathrm{max}}$}
                \State
                \State Set $i=0$, $\sigma_0 = \infty$, $h_0\left(\bm{u}\right) = \varphi_d\left(\bm{u}\right)$
                \State Sample $\mathbf{U}_0 = \{\bm{u}^k, k = 1,\dots,n\} \in \mathbb{R}^{n \times d}$
                \Comment{$\bm{u}^k \stackrel{i.i.d.}{\sim} h_0\left(\bm{u}\right)$}
                \State Compute $\mathbf{G}_0 = G(\mathbf{U}_0) \in \mathbb{R}^{n \times 1}$
                \For{$i \gets 1,i_{\mathrm{max}}$}
                \State $i \gets i+1$
                \State Compute $\sigma_i$ according to \cref{eq:sigma_i}
                \State Compute weights $\bm{\omega}_i = \left\{\Phi\left[-\mathbf{G}_{i-1}/\sigma_i\right]/\Phi\left[-\mathbf{G}_{i-1}/\sigma_{i-1}\right], k = 1,\dots,n\right\} \in \mathbb{R}^{n \times 1}$
                \State Compute $\widehat{s}_i$ according to \cref{eq:P_m_hat}
                \State $\mathbf{U}_{i-1}$ $\gets$ draw weighted resample from $\mathbf{U}_{i-1}$  with weights $\bm{\omega}_i$ \Comment sample with replacement
                \State $\left(\mathbf{U}_{i},\bm{\mathrm{G}}_{i}\right) =$ MCMC-aCS($\mathbf{U}_{i-1},\bm{\mathrm{G}}_{i-1}$,$b$)
                \Comment Details on MCMC-aCS in \cite{Papaioannou2016}
                \If{\cref{eq:exit_crit}}
                \State \textbf{break}
                \EndIf
                \EndFor
                \State Set $M \gets i$
                \State Estimate failure probability $\widehat{p}_{\mathrm{SIS}} = (\prod_{i=1}^{M}\widehat{s}_i) \frac{1}{n} \sum_{k=1}^{n}\frac{\mathrm{I}\left(\mathrm{G}_M^k \le 0\right)}{\Phi\left(-\mathrm{G}_M^k/\sigma_M\right)}$
                \Comment \cref{eq:pf_sis}
                \State \Return{$\mathbf{U}_M,\mathbf{G}_M,\widehat{p}_{\mathrm{SIS}}$.}
        \end{algorithmic}
\end{algorithm}

\section{Partial least squares--based PCEs}
\label{sec:plspce}

\subsection{PCEs}
PCEs are a tool for forward modeling the relationship between an input $\bm{X}$ and an output $Y = \mathcal{Y}(\bm{X})$.
With $\mathcal{H}$, we denote the Hilbert space of functions that are square-integrable with respect to $f_{\bm{X}}$, i.e., $\{v: \mathbb{E}_{f_{\bm{X}}} [v(\bm{X})^2] < \infty\}$.
$\mathcal{H}$ admits an inner product of two functions $v,w \in \mathcal{H}$:
\begin{equation}
\label{L2dot}
\langle v, w \rangle_{\mathcal{H}}
= \mathbb{E}_{f_{\bm{X}}(\bm{x})} [v(\bm{X}) w(\bm{X})]
= \int_{\mathbb{R}^d} v(\bm{x}) w(\bm{x}) f_{\bm{X}}(\bm{x}) \mathrm{d}\bm{x}.
\end{equation}
Let $\{v_j(\bm{X}), j\in \mathbb{N} \}$ be a complete and orthonormal basis of $\mathcal{H}$ so that $\langle v_j, v_\ell \rangle_{\mathcal{H}} = \delta_{j\ell}$, and let $\mathcal{Y} \in \mathcal{H}$. Then,
\begin{equation}
\label{eq:L2lincomb}
\mathcal{Y}(\bm{X}) = \sum_{j = 0}^{\infty} b_j v_j(\bm{X}),
\end{equation}
where the coefficients $b_j$ are defined by projecting $\mathcal{Y}$ on the basis:
\begin{equation}
\label{eq:coeffs}
b_j  = \langle\mathcal{Y}, v_j\rangle_{\mathcal{H}},~~j \in \mathbb{N}.
\end{equation}
Since $\mathcal{Y} \in \mathcal{H}$, the truncation
\begin{equation}
\label{eq:Approx}
\widehat{\mathcal{Y}}_n(\bm{X})  = \sum_{j = 0}^{n} b_j v_j(\bm{X})
\end{equation}
asymptotically converges to $\mathcal{Y}$ as $n \rightarrow \infty$ in the mean square sense.
\cite{Xiu2002} demonstrates how to construct complete orthonormal bases of $\mathcal{H}$ as polynomial families for various standard input distribution types. In particular,  if $F_{\bm{X}}(\bm{x}) = \Phi_d(\bm{x})$, where $\Phi_d$ denotes the $d$-variate independent standard-normal CDF, the tensorized, normalized probabilist's Hermite polynomials
\begin{equation}
\label{eq:HermitePol}
\Psi_{\bm{k}} (\bm{U}) = \prod_{i=1}^{d} \psi_{k_j}(U_j)
\end{equation}
form a complete orthonormal basis of $\mathcal{H}$.
$\{ \psi_j(U), j \in \mathbb{N} \}$ are the univariate, normalized (probabilist's) Hermite polynomials, and $\bm{k}=(k_1, \ldots,k_d) \in \mathbb{N}^d$.
By means of the isoprobabilistic transformation $T: \bm{X} \rightarrow \bm{U}$ introduced in the previous section, we define PCEs in standard-normal space for the remainder of the paper.
The PCE of maximum total order $p$ reads
\begin{equation}
\label{eq:pceApprox}
\widehat{\mathcal{Y}}_p(\bm{U})  = \sum_{\vert \bm{k} \vert \le p} b_{ \bm{k} } \Psi_{\bm{k}} (\bm{U}).
\end{equation}
The total number of basis functions in the PCE, $P$, depends on the input dimension $d$ and the maximum total polynomial order $p$:
\begin{equation}
\label{e:P}
P= {d + p \choose p}.
\end{equation}
The projection in \cref{eq:coeffs} can be transformed into an equivalent ordinary least squares problem \cite{Berveiller2006}.
PCEs become computationally intractable if $d$ is large, i.e., they cannot be used for problems with high-dimensional input due to the sheer number of basis functions and corresponding coefficients. In particular, the computation is rendered infeasible by the necessary number of operations to compute the set of $P$ multi-indices and the necessary number of model evaluations to obtain meaningful estimates of the coefficients.
Solution strategies to overcome these limitations (at least partially) include a hyperbolic truncation of the index set (this means to replace the condition on the $\ell_1$-norm  in \cref{eq:pceApprox}, $\vert \bm{k} \vert \le p$, with one on a general $\ell_q$-norm of $\vert \bm{k} \vert_{\alpha} = (\sum_{i=1}^d p_i^q)^{1/q}\le p$ with $q < 1$) or enforcing a maximum interaction order (i.e., a maximum number of nonzero entries in $\bm{k}$) \cite{Blatman2008}.
These approaches result in more parsimonious models and allow for PCEs to be applied in higher-dimensional problems; however, they do so at the cost of decreased model expressivity.
Sparsity-inducing solvers have been proposed to relax the dimensionality constraint imposed by the size of the regression problem. Approaches may be based on a variety of solvers for the $\ell_1$-regularized least squares problem such as least-angle regression that is used for PCEs in \cite{Blatman2011}, compressive sensing \cite{Yan2012}, and orthogonal matching pursuit \cite{Pati1993,Tropp2007,Doostan2011} as well as sparse Bayesian learning methods \cite{Tipping2001,Ji2008,Sargsyan2014,Tsilifis2020}.
For a comprehensive overview, the reader is referred to the recent literature review and benchmark study \cite{Luethen2021,Luethen2021b}.\enlargethispage*{1pc}
%%%%%%%%%%%%%%%%%%%%%%%%%%%%%%%%%%%%%%%%%%%%%

\subsection{Basis adaptation via partial least squares}
In order to obtain a parsimonious yet expressive model, we turn to low-dimensional model representations rather than sparse solutions to the full-dimensional model. To achieve this, the PCE representation is rotated onto a new basis defined by the variables $\bm{Z} = \mathbf{Q}^\mathrm{T} \bm{U}$, where $\mathbf{Q} \in \mathbb{R}^{d \times d}$ and $\mathbf{Q}^\mathrm{T} \mathbf{Q} = \mathbf{I}$, with $\mathbf{I}$ denoting the identity matrix.
This was first proposed in \cite{Tipireddy2014}. The PCE with respect to the novel basis reads
\begin{equation}
\label{eq:pceApproxQ}
\widehat{\mathcal{Y}}_p^{\mathbf{Q}}(\bm{U}) = \sum_{\vert \bm{k} \vert \le p} a_{ \bm{k} } \Psi_{\bm{k}} (\bm{Z})= \sum_{\vert \bm{k} \vert \le p} a_{ \bm{k} } \Psi_{\bm{k}} \left( \mathbf{Q}^\mathrm{T} \bm{U} \right).
\end{equation}
With $\bm{U}$ a standard-normal random vector and $\mathbf{Q}$ an orthogonal matrix, $\bm{Z}$ is a standard-normal random vector. Therefore, both original and transformed input space possess the same PCE basis, namely, the probabilist's Hermite polynomials. Merely, a new set of coefficients $a_{ \bm{k} }$ enters the formulation in the adapted basis.
The columns of $\mathbf{Q}$ define linear combinations of the original input. We seek to choose $\mathbf{Q}$ such that most of the relevant information to construct an accurate surrogate $\mathcal{Y}$ is captured in the first $m$ directions, where $m < d$ leads to dimensionality reduction.
We retain only these first $m$ columns of $\mathbf{Q}$ in the matrix $\mathbf{Q}_m$ and define a corresponding PCE of reduced dimension as
\begin{equation}
\label{eq:pceApproxQm}
\widehat{\mathcal{Y}}_p^{\mathbf{Q}_m}(\bm{U}) = \sum_{\vert \bm{k} \vert \le p} a_{ \bm{k} } \Psi_{\bm{k}} \left( \mathbf{Q}_m^\mathrm{T} \bm{U} \right),
\end{equation}
where $\bm{k} \in \mathbb{N}^m $. \cite{Tipireddy2014} computes the basis adaptation $\mathbf{Q}_m$ by evaluating first- or second-order PCE coefficients only with a sparse-grid numerical quadrature. \cite{Tsilifis2019} couples this approach with compressive sensing to simultaneously identify  $\mathbf{Q}_m$ and the PCE coefficients in the subspace. In \cite{Papaioannou2019}, we show that important directions can be identified efficiently based on a set of original function evaluations via partial least squares (PLS).

PLS establishes a linear relationship between variables $\bm{U}$ and $Y$ based on $n_{\mathcal{E}}$ observations of both quantities \cite{Wold1984}. By $\mathbf{U}_{\mathcal{E}} \in \mathbb{R}^{n_{\mathcal{E}} \times d}$, we denote the matrix of $n_{\mathcal{E}}$ observations of $\bm{U}$, and by $\mathbf{Y}_{\mathcal{E}} \in \mathbb{R}^{n_{\mathcal{E}} \times 1}$ we denote the corresponding vector of scalar responses. PLS sequentially identifies $m$ latent components $\{\bm{t}_j\}_{j=1}^m$, where $\bm{t}_j \in \mathbb{R}^{n_{\mathcal{E}} \times 1}$ such that they have maximum covariance with $\mathbf{Y}_{\mathcal{E}}$. After determining each $\bm{t}_j$, PLS assumes a linear relationship between $\bm{t}_j$ and $\mathbf{Y}_{\mathcal{E}}$ and evaluates the corresponding coefficient $a_j$ of $\bm{t}_j$ by ordinary least squares.
After each iteration, the matrices $\mathbf{U}_{\mathcal{E}}$ and $\mathbf{Y}_{\mathcal{E}}$ are deflated by the contribution of the $j$th PLS component. Components are extracted until a certain error criterion is met, which can be formulated, e.g., through the norm of the residual response vector or via cross-validation.
%%%\added[id=2]
{Dimensionality-reducing regression methods such as PLS-based regression are known to shrink the regression coefficients towards zero to produce biased estimates in exchange for reducing the estimator variances (bias-variance tradeoff). In this way, these dimensionality-reducing methods are able to produce smaller overall mean squared estimation errors (see, e.g., \cite{DeJong1995} for PLS).}

The nonlinear version of PLS in turn relaxes the assumption of a linear relationship between latent component and the response. A number of nonlinear PLS algorithms have been proposed \cite{Rosipal2010}. Here we employ the approach of  \cite{Wold1989,Baffi1999} that introduces an additional loop into the algorithm for running a Newton--Raphson procedure iterating between the current latent component and the response.
Ultimately, we are interested in computing the orthogonal transformation matrix $\bm{Q}_m$ in \cref{eq:pceApproxQm}. PLS produces two different matrices $\bm{R}$ and $\bm{W}$ that are suitable to this end, which motivates two different flavors of PLS-PCE.
In PLS-PCE-R as proposed in \cite{Papaioannou2019} (see subsection~\ref{sec:pls_pce_r}), each nonlinear relationship between the $\{\bm{t}_j\}_{j=1}^m$ and the response is modeled as a univariate PCE. The coefficients of these univarate PCEs are computed simultaneously with the latent structure, and the resulting model is a sum of univariate PCEs.
Alternatively, the univariate PCEs are discarded after the PLS-PCE algorithm terminates, and a multivariate (sparse) PCE is constructed in the subspace formed by the so-called weights $\{\bm{w}_j\}_{j=1}^m$ leading to PLS-PCE-W (see subsection~\cref{sec:pls_pce_w}).
%%%%%%%%%%%%%%%%%%%%%%%%%%%%%%%%%%%%%%%%%%%%%

% Figure 2
\begin{figure}[t!]
\vspace*{-21pt}
        \centering
        \subfloat[PLS-PCE-R]{\includegraphics[height=0.5\textwidth]{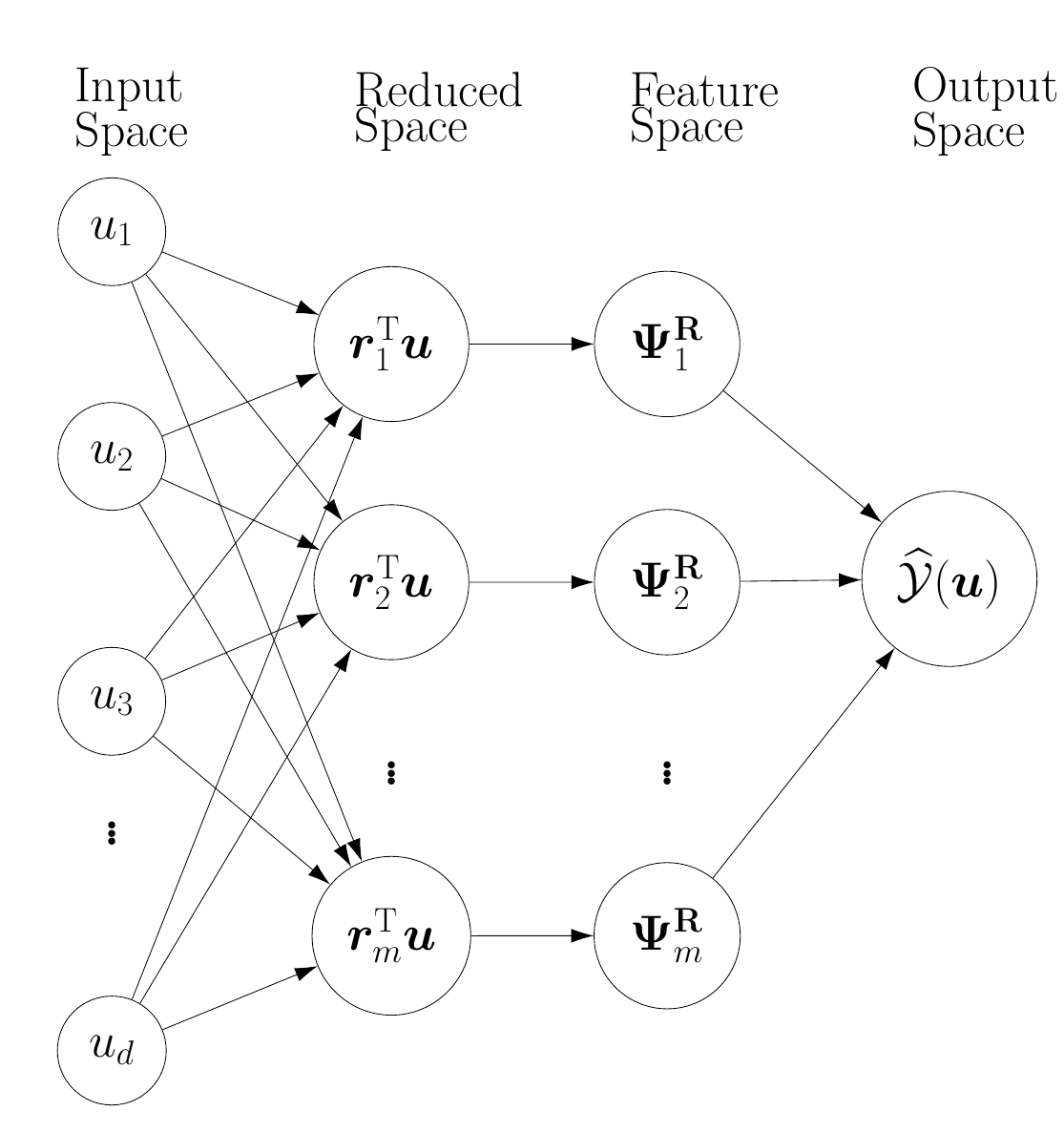}}\hfill
        \subfloat[PLS-PCE-W]{\raisebox{0px}{\includegraphics[height=0.5\textwidth]{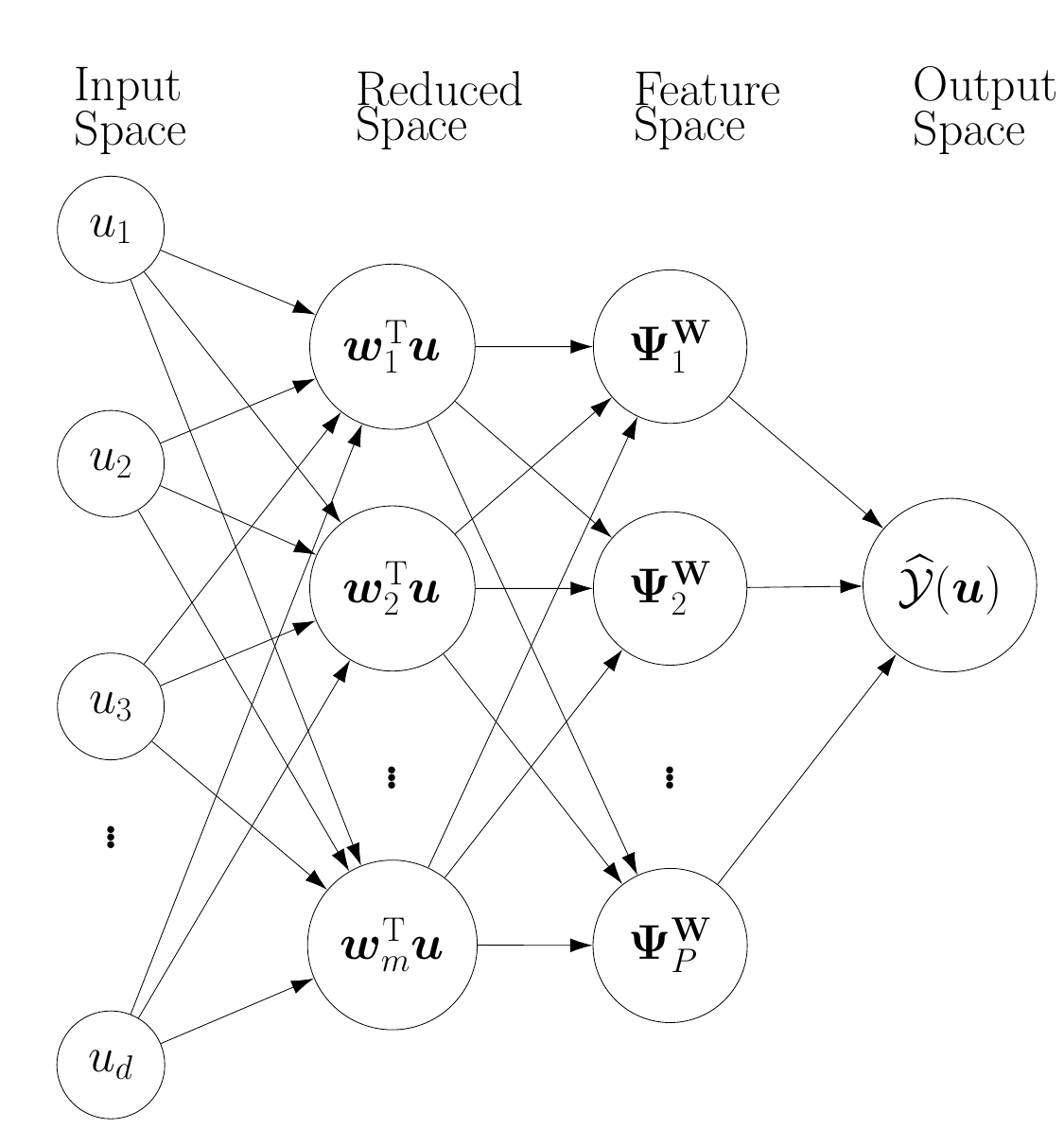}}}
        \caption{Structure of two different PLS-PCE models, where $\bm{\Psi}_j^{\mathbf{W}} = \bm{\Psi}_{\bm{\alpha}_j}$ as defined in \cref{eq:PLS_PCE_W} and  $\bm{\Psi}_j^{\mathbf{R}} = (\widehat{\bm{a}}_j^{q_j})^{\mathrm{T}}\bm{\psi}_{q_j}$ as seen from \cref{eq:PLS_PCE_R}. Essential differences exist in the choice of the reduced space basis (layer $2$) and the modeling of cross-terms when mapping from reduced to feature space (layers $2$ \& $3$) with PLS-PCE-W (b).}
        \label{fig:pls_pce_structure}
\end{figure}

\subsection{PLS-PCE-R}\label{sec:pls_pce_r}
PLS-PCE-R identifies $m$ latent components, and for each component, it returns the direction $\bm{r}_j$ and the univariate PCE along this direction. The univariate PCEs are defined by their polynomial orders $\{q_j\}_{j=1}^m$ and the associated coefficient vectors $\{\bm{a}_j\}_{j=1}^m$. The polynomial order is identified with leave-one-out cross validation \cite{Chapelle2002}. For each ($j$th) latent component, the nonlinear PLS iteration is repeated for different polynomial orders, and $q_j$ is chosen as the order minimizing the leave-one-out error.
The PLS-PCE-R model reads
\begin{equation}
\label{eq:PLS_PCE_R}
%\widehat{Y}_m^{\mathrm{PLS}}=b_0 + \sum_{i=1}^m \sum_{j=1}^{q_j} (\bm{a}_{i})_j \psi_{j}\left[ (\bm{r}_j)^\mathrm{T} \tilde{\bm{U}}) \right],
\widehat{\mathcal{Y}}(\bm{u})=\widehat{a}_0 + \sum_{j=1}^m \left(\widehat{\bm{a}}_j^{q_j}\right)^\mathrm{T} \bm{\psi}_{q_j}\left[ \bm{r}_j^\mathrm{T} \left(\bm{u} -  \bm{\mu}_{\mathbf{U}}\right) \right],
\end{equation}
where $\widehat{a}_0 = \widehat{\mathbb{E}}[\mathbf{Y}]$, $\bm{\psi}_{q_j}(\bm{U})$ is a vector function assembling the evaluations of the one-dimensional Hermite polynomials up to order $q_j$, and $\bm{\mu}_{\mathbf{U}}$ is the columnwise sample mean of $\mathbf{U}_{\mathcal{E}}$.
The model structure is illustrated in Figure~\ref{fig:pls_pce_structure}.
The PLS directions $\bm{r}_j$ can be evaluated in terms of the PLS weights $\bm{w}_j$ and loads $\bm{p}_j$ through the following recursive relation \cite{Hoeskuldsson1988}:
\begin{equation}
\label{eq:PLSRmatrixRecursive}
\begin{aligned}
\bm{r}_1 &= \bm{w}_1, \\
\bm{r}_j &= \bm{w}_j-\bm{r}_{j-1}\left(\bm{p}^\mathrm{T}_{j-1} \bm{w}_j\right).
\end{aligned}
\end{equation}
$\mathbf{R} = [\bm{r}_1, \ldots, \bm{r}_m] \in \mathbb{R}^{d \times m}$ is a matrix collecting all PLS directions. $\mathbf{R}$ is not necessarily orthogonal; i.e., in general $\mathbf{R}^\mathrm{T} \mathbf{R} \ne \mathbf{I}$.
However, in \cite{Papaioannou2019} it is shown that $\mathbf{R}^\mathrm{T} \mathbf{R} \approx \mathbf{I}$ when $n_{\mathcal{E}}$ is large and $\mathbf{U}_{\mathcal{E}}^\mathrm{T} \mathbf{U}_{\mathcal{E}}$ is diagonal, which is the case if $\mathbf{U}_{\mathcal{E}}$ is drawn from $\varphi_d$. In this case, \cref{eq:PLS_PCE_R} is equivalent to a PCE of the form \cref{eq:pceApproxQm}, where only main effects in the latent components are considered.
%%%%%%%%%%%%%%%%%%%%%%%%%%%%%%%%%%%%%%%%%%%%%

\subsection{PLS-PCE-W}\label{sec:pls_pce_w}
PLS-PCE-W defines $\mathbf{W}$ as basis of the subspace rather than $\mathbf{R}$, where $\mathbf{W} = [\bm{w}_1, \ldots, \bm{w}_m] \in \mathbb{R}^{d \times m}$.
Within linear PLS, the columns of $\mathbf{W}$ form an orthogonal basis. Within nonlinear PLS, the Newton--Raphson step may introduce deviations from orthogonality, which are however negligible in all tested examples.
The univariate PCEs obtained through the Newton--Raphson step will be optimal with respect to $\mathbf{R}$, not $\mathbf{W}$. Thus, in PLS-PCE-W these univariate polynomials are discarded once $\mathbf{W}$ is identified and a multivariate (sparse) PCE is constructed in the subspace defined by $\mathbf{W}$ using least-angle regression and a hyperbolic truncation scheme for the multivariate PCE basis as proposed by \cite{Blatman2011}. In this way PLS-PCE-W achieves more flexibility compared to PLS-PCE-R by including interactions of the latent components in exchange for a departure from optimality in the match between latent component and surrogate model.
In analogy to \cref{eq:pceApproxQm}, the PLS-PCE-W model reads
\begin{equation}
\label{eq:PLS_PCE_W}
\widehat{\mathcal{Y}}(\bm{u})=\widehat{a}_0 + \sum_{\bm{k} \in \bm{\alpha}} \widehat{a}_{\bm{k}}  \bm{\Psi}_{\bm{k}} \left[ \mathbf{W}^\mathrm{T} \left(\bm{u} -  \bm{\mu}_{\mathbf{U}}\right) \right],
\end{equation}
where $\bm{\alpha} \in \mathbb{N}^{P \times d}$ is the multi-index set, which indicates the polynomial orders of the $d$ univariate polynomials in each of the $P$ multivariate polynomials as obtained with least-angle regression. Both PLS-PCE-R and PLS-PCE-W are summarized in \cref{alg:PLS-PCE}. In the following, we will use the PLS-PCE-W model, as we observed a superior performance for this model compared to PLS-PCE-R models in the context of the proposed approach.

\begin{algorithm}
        \caption{PCE-driven PLS algorithm \cite{Papaioannou2019}} \label{alg:PLS-PCE}
        \begin{algorithmic}[1]
                \State \textbf{Input} Input matrix $\mathbf{U}_{\mathcal{E}}$ and output vector $\mathbf{Y}_{\mathcal{E}}$, maximum polynomial order~$p$
                \State
                \State Set $\mathbf{E}=\mathbf{U}_{\mathcal{E}} - \bm{\mu}_{\mathbf{U}}$, $\mathbf{F}=\mathbf{Y}_{\mathcal{E}} - \bm{\mu}_{\mathbf{Y}}$, $\epsilon_w = 10^{-3}$, $\epsilon_y = 10^{-3}$, $j=1$
                \Repeat
                \State Compute weight $\bm{w}_j^0  =\mathbf{E}^\mathrm{T}\mathbf{F} / \Vert \mathbf{E}^\mathrm{T}\mathbf{F} \Vert$
                \For{$q \gets 1,p$}
                \State Set $\bm{w}_j^q = \bm{w}_j^0$
                \Repeat
                \State Compute score $\bm{t}_j^q  =\mathbf{E}\bm{w}_j^q$
                \State Fit a 1D PCE of order $q$ $\widehat{\bm{a}}^q_j \gets \mathrm{fit} \left[ \mathbf{F} = (\bm{a}^q_j)^\mathrm{T} \bm{\psi}_{q}(\bm{t}_j^q) + \bm{\epsilon}\right]$
                \State Set $\widehat{\mathcal{M}}_j^q(t) = (\widehat{\bm{a}}^q_j)^\mathrm{T} \bm{\psi}_{q}(\bm{t}_j^q)(t)$
                \State Compute the error $\bm{e} = \mathbf{F}-(\widehat{\bm{a}}^q_j)^T \bm{\psi}_{q}(\bm{t}_j^q)$
                \State Compute $\Delta \bm{w}_j^q = (\mathbf{A}^\mathrm{T}\mathbf{A})^{-1}\mathbf{A}^\mathrm{T}\bm{e}$ with $\mathbf{A}=\nabla_{\bm{w}} (\widehat{\bm{a}}^q_j)^\mathrm{T} \bm{\psi}_{q} (\mathbf{E}\bm{w})$
                \State Set $\bm{w}_j^q \gets  \bm{w}_j^q + \Delta \bm{w}_j^q$
                \State Normalize $\bm{w}_j^q \gets \bm{w}_j^q  / \Vert \bm{w}_j^q \Vert$
                \Until{$\Vert\Delta \bm{w}_j^q\Vert$ is smaller than $\epsilon_w$}
                \State Evaluate the relative leave-one-out error $\epsilon_{LOO}^q$ as in \cite{Blatman2011}
                \EndFor
                \State Set $\{q_j,\widehat{\bm{a}}_j^{q_j},\bm{w}_j^{q_j}\}$ as the triple $\{q,\widehat{\bm{a}}_j^{q},\bm{w}_j^{q}\}$ with the smallest $\epsilon_{LOO}^q$
                \State Compute score: $\bm{t}_j^{q_j}  =\mathbf{E}\bm{w}_j^{q_j}$
                \State Compute load: $\bm{p}_j^{q_j}  =\mathbf{E}^\mathrm{T} \bm{t}_j^{q_j} / ((\bm{t}_j^{q_j})^\mathrm{T} \bm{t}_j^{q_j})$
                \State Deflate: $\mathbf{E} \gets \mathbf{E}-\bm{t}_j^{q_j} (\bm{p}_j^{q_j})^\mathrm{T}$, $\mathbf{F} \gets \mathbf{F}-  (\widehat{\bm{a}}^{q_j}_j)^\mathrm{T} \bm{\psi}_{q_j}(\bm{t}_j^{q_j})$
                \State $j \gets j+1$
                \Until{change in $\Vert \mathbf{F} \Vert$ is smaller than $\epsilon_y$}
                \State Compute $\mathbf{R} = [\mathbf{r}_1,\mathbf{r}_2,\dots,\mathbf{r}_m]$ according to \cref{eq:PLSRmatrixRecursive} \Comment For the $R$-based version of PLS-PCE
                \State Build $\widehat{\mathcal{Y}}(\bm{u})$ according to \cref{eq:PLS_PCE_R}
                \State Gather $\mathbf{W} = [\bm{w}_1,\bm{w}_2,\dots,\bm{w}_m]$
                \Comment For the $W$-based version of PLS-PCE
                \State Build $\widehat{\mathcal{Y}}(\bm{u})$ according to \cref{eq:PLS_PCE_W} and \cite{Blatman2011}
                \State \Return{$\mathbf{R}$/$\mathbf{W}$, $\widehat{\mathcal{Y}}(\bm{u})$}
        \end{algorithmic}
\end{algorithm}

%%%%%%%%%%%%%%%%%%%%%%%%%%%%%%%%%%%%%%%%%%%%%
\section{Learning PLS-PCE models in each SIS level}\enlargethispage*{0.2pc}
%%%%%%%%%%%%%%%%%%%%%%%%%%%%%%%%%%%%%%%%%%%%%

\subsection{The sequential subspace importance sampler}
We recently proposed to reconstruct low-dimensional PLS-PCE-W models in each level of SIS to improve the tractability of high-dimensional reliability analysis with computationally expensive models \cite{Papaioannou2018b}. We term this approach sequential subspace IS or SSIS.
The efficiency of SIS benefits from surrogate modeling through a considerable reduction of required\pagebreak{} model evaluations. The PLS-PCE model alone, being a global surrogate model, is a relatively limited tool for reliability analysis. Combining it with SIS provides the means to sequentially move the training set towards relevant regions in the input space and thereby renders difficult reliability problems accessible to surrogate modeling.
At the $i$th SSIS level, a new \textit{local} training set is sampled
from the current importance density $h_i$ through a resampling step on the $N$ available samples from $h_i$. The new \textit{local} training set is appended to the \textit{global} training set comprising earlier designs from levels $1$ through $i-1$. Based on the updated \textit{global} training set, a new PLS-PCE model is constructed and SIS is rerun for $i+1$ levels from $h_0$ to obtain samples for the next \textit{local} training set. Due to this restart, it is sensible to let previously used local training sets remain in the global training set such that the $i$th surrogate model accurately predicts the LSF output along the entire path of samples moving from the nominal distribution $h_0$ to $h_i$.
The restart itself incurs no additional LSF evaluations and serves to stabilize the method: Without restart, the computation of $\sigma_{i+1}$ according to \cref{eq:sigma_i} is based on two different surrogate models: the most recent model constructed in level $i$ appears in the numerator of the sample CoV of the weights and the model constructed in level $i-1$ appears in the denominator. These models may however be too different from one another to admit a solution in \cref{eq:sigma_i}, i.e., to achieve the prescribed CoV $\delta_{\text{target}}$ between two subsequent IS densities.

In an additional step, before propagating the intermediate importance density to the next level of the SSIS algorithm, we introduce AL. This ensures a prescribed surrogate model accuracy in regions of high probability mass of the current sampling density. In turn, this refined surrogate model is used to propagate samples to the next level.
When the underlying SIS algorithm reaches convergence, a final AL procedure, performed over samples of the final importance density, ensures that the probability of failure is estimated with a surrogate model that captures the failure hypersurface well.
This approach is termed adaptive SSIS or ASSIS.

AL emerged in the late 1980s as a subfield of machine learning \cite{Settles2009} and has been known in the statistical theory of regression as optimal experimental design since the early 1970s \cite{Fedorov1972}. At its heart is the idea that supervised learning algorithms can perform better if allowed to choose their training data.
We consider a ``pool-based sampling'' variant of AL, in which a large pool of unlabeled data points is made available to the algorithm. Within SIS, one has $n$ samples from $h_i$ available in the $i$th level. The algorithm then selects $n_{\mathrm{add}}$ points that are labeled (i.e., for which the LSF is evaluated) and added to the training set based on a measure of information gain. This measure typically takes the form of a learning function $\mathcal{L}$ that is maximized over the sample pool to perform selection.
The learning function employed in the context of SSIS is discussed in subsection~\ref{sec:AL}.

\begin{figure}[t!]
        \centering
        \subfloat[SSIS]{\includegraphics[width=0.43\textwidth]{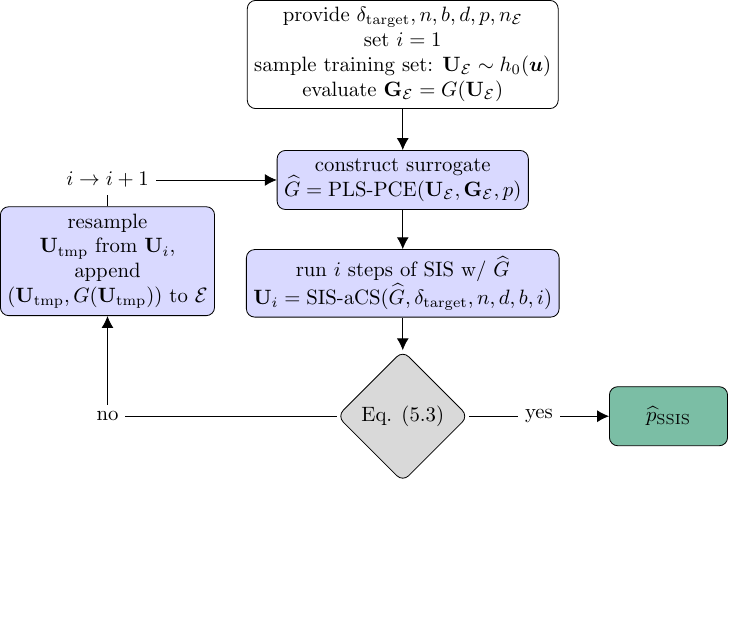}}\hfill
        \subfloat[ASSIS]{\raisebox{-0px}{\includegraphics[width=0.53\textwidth]{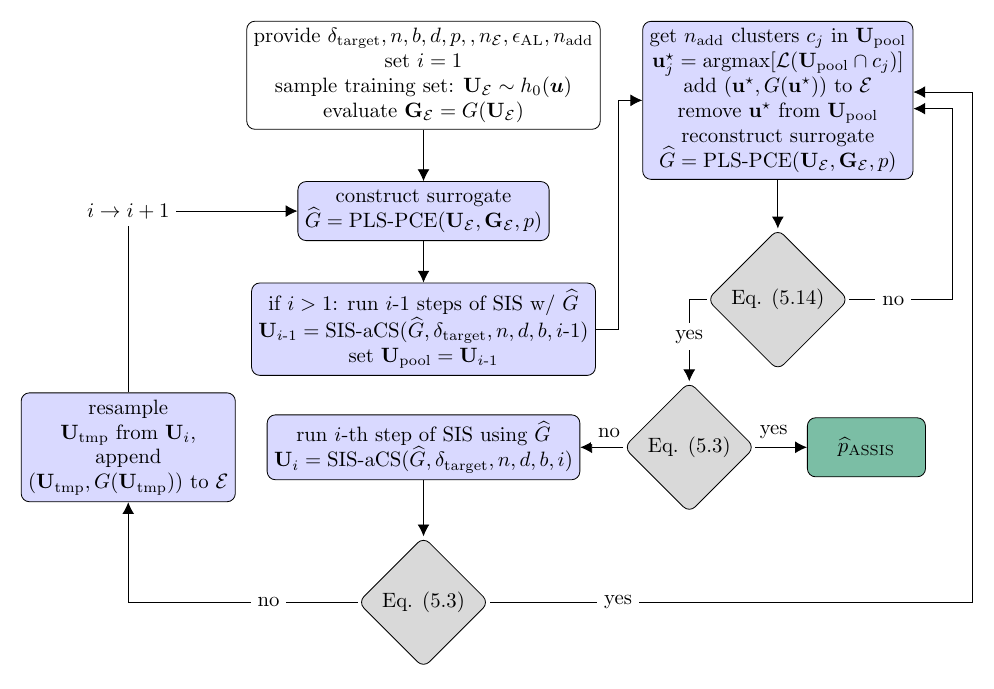}}}
        \caption{Comparison of SIS-PLS-PCE with (right) and without (left) AL.}
        \label{fig:schematic}\vspace*{-8pt}
\end{figure}

The probability of failure estimator for SSIS/ASSIS is analogous to \cref{eq:pf_sis} with the difference that SIS is performed with an LSF approximation $\widehat{G}$ that is based on the final surrogate model:
\begin{equation}
\label{eq:pf_ssis}
\widehat{p}
= \left(\prod_{i=1}^M \widehat{s}_i\right) \frac{1}{n}\sum_{k=1}^{n} \frac{\mathrm{I}(\widehat{G}\left(\bm{u}^k\right) \le 0) \varphi_d\left(\bm{u}^k\right)}{\eta_M\left(\bm{u}^k\right)},~~~\bm{u}^k \stackrel{i.i.d.}{\sim} h_M.
\end{equation}
The ratio of normalizing constants $\{\widehat{s}_i\}_{i=1}^M$ is estimated as
\begin{equation}
\label{eq:pf_ssis_aux}
\widehat{s}_i
= \frac{1}{n}\sum_{k=1}^{n} \widehat{\omega}_i\left(\bm{u}^k\right)
= \frac{1}{n}\sum_{k=1}^{n} \frac{\Phi(-\widehat{G}\left(\bm{u}^k\right)/\sigma_i)}{\Phi(-\widehat{G}\left(\bm{u}^k\right)/\sigma_{i-1})},~~~\bm{u}^k \stackrel{i.i.d.}{\sim} h_i.
\end{equation}
The SSIS/ASSIS algorithms are stopped based on a similar criterion as for SIS given in \cref{eq:exit_crit}:
\begin{equation}
\label{eq:exit_crit_ssis_al}
\widehat{\mathbb{COV}}\left[\frac{\mathrm{I}(\widehat{G}(\bm{U}) \le 0)}{\Phi(-\widehat{G}\left(\bm{U}\right)/\sigma_i)}\right] \le \delta_{\mathrm{target}}.
\end{equation}
Figure~\ref{fig:schematic} depicts flow diagrams of the SSIS and ASSIS algorithms.

% Figure 3

%%%%%%%%%%%%%%%%%%%%%%%%%%%%%%%%%%%%%%%%%%%%%
\subsection{AL of low-dimensional model representations}
\label{sec:AL}
In the context of SSIS, the learning function $\mathcal{L}$ should express the prediction uncertainty at each sample of the current IS density for a given PLS-PCE-W surrogate. This prediction uncertainty is due to the estimation of both the subspace and the surrogate model with a finite-sized training set.
We describe this uncertainty with the variance of the LSF based on the surrogate model conditional on $\bm{u}$,  $\mathbb{V}[\widehat{G}|\bm{U} = \bm{u}]$. Note that, whenever the distribution with respect to which $\mathbb{E}[\cdot]$ or $\mathbb{V}[\cdot]$ is evaluated is not made explicit as a subscript, it is implicitly assumed as the distribution of the argument. For example, $\mathbb{V}[\widehat{G}|\bm{U} = \bm{u}] = \mathbb{V}_{f_{\widehat{G}|\bm{u}}}[\widehat{G}|\bm{U} = \bm{u}]$.

Let $\bm{\xi}_0 = \bm{a} \in \mathbb{R}^{P \times 1}$ and $\bm{\xi}_j = \bm{w}_j \in \mathbb{R}^{d \times 1},~j=1,\dots,m$, such that $\bm{\xi} = [\bm{\xi}_0^{\mathrm{T}}, \bm{\xi}_1^{\mathrm{T}} \dots, \bm{\xi}_m^{\mathrm{T}}]^{\mathrm{T}}  \in \mathbb{R}^{(md + P) \times 1}$ is the collection of all $md + P$ model parameters.
Further, let $\bm{\xi}^\star$ denote their corresponding point estimates returned by \cref{alg:PLS-PCE}.
The first-order expansion of $\widehat{\mathbb{V}}[\widehat{G}|\bm{u}]$ around $\bm{\xi}^\star$ reads
\begin{equation}
\label{eq:var_G}
\widehat{\sigma}^2_{\widehat{G}}(\bm{u}) = \widehat{\mathbb{V}}[\widehat{G}|\bm{u}] \approx \left[\frac{\partial \widehat{G}}{\partial\bm{\xi}}\right]^{\mathrm{T}}_{\bm{\xi} = \bm{\xi}^\star}  \widehat{\bm{\Sigma}}_{\bm{\xi}\bm{\xi}} \left[\frac{\partial \widehat{G}}{\partial\bm{\xi}}\right]_{\bm{\xi} = \bm{\xi}^\star},
\end{equation}
where $\widehat{\bm{\Sigma}}_{\bm{\xi}\bm{\xi}}$ is an estimate of the parameter covariance matrix. Next, we neglect the pairwise cross-covariance of PCE coefficients $\bm{a}$ and the subspace components $\bm{w}_j$ and consider
\begin{equation}
\label{eq:var_G_componentwise}
\widehat{\sigma}^2_{\widehat{G}}(\bm{u}) =\widehat{\mathbb{V}}[\widehat{G}|\bm{u}] \approx
\sum_{j=0}^m \left[\frac{\partial \widehat{G}(\bm{u},\bm{\xi})}{\partial\bm{\xi}_j}\right]^{\mathrm{T}}_{\bm{\xi}_j = \bm{\xi}_j^\star} \widehat{\bm{\Sigma}}_{\bm{\xi}_j\bm{\xi}_j} \left[\frac{\partial \widehat{G}(\bm{u},\bm{\xi})}{\partial\bm{\xi}_j}\right]_{\bm{\xi}_j = \bm{\xi}^\star}.
\end{equation}
This significantly reduces the number of $\bm{\Sigma}_{\bm{\xi}\bm{\xi}}$-entries that have to be estimated, namely, from $P^2 + 2 P m d +m^2 d^2$ to $P^2+m d^2$. More importantly, the coefficients of the PCE, $\bm{\xi}_0$, are obtained with linear regression while the subspace, $\{\bm{\xi}_j\}_{j=1}^{m}$, is obtained in the inner loop of \cref{alg:PLS-PCE} with nonlinear regression.
%%%\added[id=1]
{Due to this sequential estimation of the \smash{$\{\bm{\xi}_j\}_{j=0}^m$}, there is no straightforward way of obtaining an estimate of the full covariance matrix. In particular, we are not aware of such an estimate for the parameters of nonlinear PLS. Hence, this simplification is not only convenient but also necessary in practice. We do observe, however, that the off-diagonal elements of the estimated componentwise cross-covariance matrices \smash{$\widehat{\bm{\Sigma}}_{\bm{\xi}_j\bm{\xi}_j}$} are several orders of magnitude smaller compared to the the main diagonal elements. This indicates that the model uncertainty estimate is dominated by parameter variances. In fact, in a more radical approach that remains unexplored in this work, one may consider parameter variances only (i.e., only $P + m d$ entries of the full covariance matrix are retained). Such an approach is, e.g., used in \cite{Perrin2020}.}
Under some regularity conditions, the estimator $\bm{\xi}_j^*$ is consistent \cite{Wu1981} and converges in distribution to a multivariate Gaussian distribution with mean $\bm{\xi}_j$ and covariance $ \mathbf{\Sigma}_{\bm{\xi}_j\bm{\xi}_j}$. In analogy with linear regression, an estimate of $\mathbf{\Sigma}_{\bm{\xi}_j\bm{\xi}_j}$ is given through
\begin{equation}
\label{eq:xi_dist}
\widehat{\bm{\Sigma}}_{\bm{\xi}_j\bm{\xi}_j} =
%%%\added[id=3]
{\widehat{\sigma}_{\epsilon}^2} \left(\mathbf{A}_j^{\mathrm{T}}\mathbf{A}_j\right)^{-1}
\end{equation}
with
\begin{equation}
\label{eq:xi_dist_aux}
\mathbf{A}_j =
\left[\frac{\partial \widehat{\mathcal{Y}}(\bm{u},\bm{\xi})}{\partial\bm{\xi}_j}\right]_{\substack{\bm{\xi} = \bm{\xi}^\star\\\bm{u}=\mathbf{U}_{\mathcal{E}}}} \in\mathbb{R}^{n_{\mathcal{E}} \times d}
~~~\mathrm{and}~~~
\widehat{\sigma}^2_{\epsilon} = \frac{1}{n_{\mathcal{E}}-md-P}\sum\limits_{k=1}^{n_{\mathcal{E}}} \left[\mathbf{Y}_{\mathcal{E}}^k-\widehat{\mathcal{Y}}(\mathbf{U}_{\mathcal{E}}^k)\right]^2.
\end{equation}
$\widehat{\sigma}^2_{\epsilon}$ is the standard estimator for the error variance of the surrogate model.
$\mathbf{A}_j$ is the gradient of the surrogate model $\mathcal{Y}$ with respect to the model parameters evaluated at each of the $n_{\mathcal{E}}$ points in the training set $\mathbf{U}_{\mathcal{E}}$. $\mathbf{A}_0$ is merely the design matrix and does not require the computation of any derivatives. Note that computing the gradients $\{\mathbf{A}_j\}_{j=0}^m$ does not require any model evaluations.
For $j=0$, it is
\begin{equation}
\frac{\partial \widehat{\mathcal{Y}}(\bm{u},\bm{\xi})}{\partial\bm{\xi}_0}
= \left[\bm{\Psi}_{i} \left(\mathbf{W}^\mathrm{T} \left(\bm{u} -  \bm{\mu}_{\mathbf{U}}\right) \right)\right]_{i=1}^{P-1}~~~\text{with}~~~\mathbf{W} = \left[\bm{\xi}_1,\bm{\xi}_2,\dots,\bm{\xi}_m\right].
\end{equation}
For $j>0$ and recalling $\bm{z} = \mathbf{W}^{\mathrm{T}}(\bm{u} -  \bm{\mu}_{\mathbf{U}})$, we have
\begin{equation}
\begin{split}
\frac{\partial\Psi_{\bm{k}} (\bm{z})}{\partial \bm{\xi}_j} &=
\frac{\partial}{\partial \bm{w}_j}\Psi_{\bm{k}} (\mathbf{W}^{\mathrm{T}}(\bm{u} -  \bm{\mu}_{\mathbf{U}}))\\
&= \left(\bm{u} -  \bm{\mu}_{\mathbf{U}}\right) \frac{\partial\Psi_{\bm{k}} (z_j)}{\partial z_j}\\
&= \left(\bm{u} -  \bm{\mu}_{\mathbf{U}}\right) \left(\prod\limits_{\substack{i=1\\i \neq j}}^m \psi_{k_i} (\bm{w}_i^{\mathrm{T}}\bm{u})\right) \frac{\partial \psi_{k_j} (\bm{w}_j^{\mathrm{T}}\bm{u}) }{\partial z_j}\\
&= \left(\bm{u} -  \bm{\mu}_{\mathbf{U}}\right) \left(\prod\limits_{\substack{i=1\\i \neq j}}^m \psi_{k_i} (\bm{w}_i^{\mathrm{T}}\bm{u})\right)\sqrt{k_j} \psi_{k_j-1} (\bm{w}_j^{\mathrm{T}}\bm{u}).
\end{split}
\end{equation}
In the last equality, we have used the following expression for derivatives of univariate normalized Hermite polynomials:
\begin{equation}
\frac{\mathrm{d}\psi_{n} (x)}{\mathrm{d} x} = \sqrt{n} \psi_{n-1} (x).
\end{equation}
$\partial \widehat{\mathcal{Y}}(\bm{u},\bm{\xi})/\partial\bm{\xi}_j$ for $j>0$ follows as
\begin{equation}
\frac{\partial \widehat{\mathcal{Y}}(\bm{u},\bm{\xi})}{\partial\bm{\xi}_j}
= \frac{\partial \widehat{\mathcal{Y}}(\bm{z})}{\partial\bm{w}_j}
= \sum_{\bm{k} \in \bm{\alpha}} \widehat{a}_{\bm{k}}  \frac{\partial\Psi_{\bm{k}} (\bm{z})}{\partial \bm{\xi}_j},~~~~~j>0.
\end{equation}
The partial derivative $\partial \widehat{G}/\partial\bm{\xi}_j$ in \cref{eq:var_G_componentwise} can be evaluated using the chain rule of differentiation, which yields
\begin{equation}
\label{eq:chain_rule_1}
\frac{\partial \widehat{G}}{\partial\bm{\xi}_j} = \frac{\partial  \widehat{G}}{\partial \widehat{\mathcal{Y}}} \frac{\partial \widehat{\mathcal{Y}}}{\partial\bm{\xi}_j}.
\end{equation}
The first term on the right-hand side is typically easy to compute and often equals $\pm1$ (the sign is irrelevant as the gradient enters the quadratic form in \cref{eq:var_G_componentwise}) if the LSF returns the difference between the model output and a prescribed threshold. In this case, the first factor on the right-hand side of \cref{eq:chain_rule_1} drops out. If, however, the LSF is not continuously differentiable with respect to the model, we may construct a surrogate model of $G$ directly by using a training set containing LSF evaluations rather than model evaluations in \cref{alg:PLS-PCE}.
The second term on the right-hand side can be obtained reusing the gradients from the $\mathbf{A}_j$ in \cref{eq:xi_dist_aux} that---in this case---are not evaluated at the training set and thus are functions of $\bm{u}$.

When setting up the learning function, there is a distinction to be made between an intermediate SIS level and the final SIS level: In the intermediate level, the goal is to accurately estimate the ratios of normalizing constants and to propagate the samples to the next level. In the final level, the goal is to build the probability of failure estimator and thus to accurately approximate the true limit-state hypersurface.
With this in mind, the learning functions for adapting the surrogate models in levels $i=1,\dots,M$, and after the final level are readily stated as
\begin{equation}
\label{eq:learning_function}
\mathcal{L}_G\left(\bm{u}\right)
= \begin{cases}
{\sigma}_{\widehat{G}}(\bm{u}), & \text{intermediate SIS level},\\
{\sigma}_{\widehat{G}}(\bm{u})/|\widehat{G}(\bm{u})|, & \text{after final SIS level}.
\end{cases}
\end{equation}
After the final level, SIS has converged and we are using samples from the final biasing density $h_M$ to refit a surrogate model that captures the failure hypersurface well.
The learning function in this case is defined in the spirit of the learning function put forward in \cite{Echard2011}. The denominator penalizes samples whose image under $\widehat{G}$ is far away from 0 assuming that therefore they are themselves far away from the failure hypersurface. Such samples are unlikely to be misclassified as safe if located in the failure domain or vice versa.
In all previous levels of SIS, there is no failure hypersurface to be approximated but only importance weights and the resulting ratio of normalizing constants. Here, the denominator in the learning function is dropped, as there is no benefit to penalizing samples with large absolute image values under $\widehat{G}$.

In each AL iteration, the pool is searched for one or several points maximizing $\mathcal{L}\left(\bm{u}\right)$. If $n_{\mathrm{add}} > 1$ new points are added per AL iteration, the current sample pool is transformed to the low-dimensional subspace defined by $\mathbf{W}$ in order to identify $n_{\mathrm{add}}$ clusters (e.g., with $k$-means). Clustering in the subspace circumvents the performance deterioration most clustering methods experience in high dimensions \cite{Kriegel2009}. The point maximizing \cref{eq:learning_function} in each cluster is added to the training set. In this way, the algorithm avoids a local concentration of the training set in a single region and is also able to handle problems with multiple disconnected failure domains as long as these are contained in the subspace.

The AL is terminated based on the maximum local standard deviation relative to the target average in the intermediate levels or based on the relative change of the probability of failure estimate after the final level:
\begin{equation}
\label{eq:termination}
\left.
\begin{cases}
\max\limits_{k = 1,\dots,n}\left(\frac{{\sigma}_{\widehat{G}}(\bm{u}_k)}{\widehat{\mathbb{E}}[\widehat{G}(\bm{U})]}\right) \le \epsilon_{\mathrm{AL}},& \text{intermediate SIS level},\\
\frac{\widehat{p} - \widehat{p}_{\mathrm{last}}}{\widehat{p}} \le \epsilon_{\mathrm{AL}}, &  \text{after final SIS level}
\end{cases}
\right\},
\end{equation}
where appropriate choices for $\epsilon_{\mathrm{AL}}$ lie in $[10^{-2},10^{-1}]$. $\widehat{p}$ and $\widehat{p}_{\mathrm{last}}$ denote the probability of failure estimate based on the current and the last training set within the AL loop. The probability of failure is estimated with a surrogate model--based run of SIS-aCS in each AL iteration. This causes no additional cost in terms of original model evaluations and ensures a reliable evaluation of the criterion even for extremely small failure probabilities.
The AL procedure is detailed in \cref{alg:AL}, and the complete method is detailed in \cref{alg:ASSIS}.
\begin{algorithm}
        \caption{Active Learning} \label{alg:AL}
        \begin{algorithmic}[1]
                \State \textbf{Input} {LSF $G\left(\bm{u}\right)$, AL error level $\epsilon_{\mathrm{AL}}$, \# of AL clusters $n_{\mathrm{add}}$, polynomial order $p$, training set $\{\mathbf{U}_{\mathcal{E}},\mathbf{G}_{\mathcal{E}}\}$, sample pool $\mathbf{U}_{\mathrm{pool}}$}
                \State
                \While{\textit{true}} \Comment AL loop
                \State Run $[\mathbf{W}, \widehat{G}] = $ PLS-PCE$(\mathbf{U}_{\mathcal{E}},\mathbf{G}_{\mathcal{E}},p,\mathrm{'W'})$
                \Comment \cref{alg:PLS-PCE}
                \If {\cref{eq:termination}}
                \State \textbf{break}
                \EndIf
                \State Identify $n_{\mathrm{add}}$ clusters among $\mathbf{U}_{\mathrm{pool}}\mathbf{W}$ \Comment Clustering performed in the subspace defined by $\mathbf{W}$
                \For{\textbf{each} cluster}
                \State $\mathbf{U}_{\mathrm{cluster}} = \{\mathbf{u} \in \mathbf{U}_{\mathrm{pool}}: \mathbf{u} \in \mathrm{cluster}\}$
                \State Evaluate $\mathbf{u}^{\star} = \mathrm{argmax}[\mathcal{L}(\mathbf{U}_{\mathrm{cluster}})]$  according to \cref{eq:var_G_componentwise,eq:xi_dist,eq:xi_dist_aux,eq:chain_rule_1,eq:learning_function}
                \State Append $\mathbf{U}_{\mathcal{E}} \gets [\mathbf{U}_{\mathcal{E}},\mathbf{u}^{\star}]$
                \State Append $\mathbf{G}_{\mathcal{E}} \gets [\mathbf{G}_{\mathcal{E}},G(\mathbf{u}^{\star})]$
                \State Remove $\mathbf{u}^{\star}$ from $\mathbf{U}_{\mathrm{pool}}$
                \EndFor
                \EndWhile
                \State \Return{$\mathbf{U}_{\mathcal{E}}$, $\mathbf{G}_{\mathcal{E}}$, $\widehat{G}$}
        \end{algorithmic}
\end{algorithm}
%%%%%%%%%%%%%%%%%%%%%%%%%%%%%%%%%%%%%%%%%%%%%
\begin{algorithm}
        \caption{ASSIS (with PLS-PCE-W)} \label{alg:ASSIS}
        \begin{algorithmic}[1]
                \State \textbf{Input} {LSF $G\left(\bm{u}\right)$, target CoV $\delta_{\mathrm{target}}$, samples per level $n$, input dimension $d$, training set size $n_{\mathcal{E}}$, AL error level $\epsilon_{\mathrm{AL}}$, \# of AL clusters $n_{\mathrm{add}}$, polynomial order~$p$\\\hspace*{1cm} }
                \State
                \State Set  $i = 0$, $\sigma_i = \infty$, $h_i\left(\bm{u}\right) = \varphi_d\left(\bm{u}\right)$
                \State Initialize $\mathbf{U}_{\mathcal{E}} = [\cdot],~\mathbf{G}_{\mathcal{E}} = [\cdot]$
                \State Sample $\mathbf{U}_0 = \{\bm{u}^k\}_{k=1}^n \in \mathbb{R}^{n \times d}$
                \Comment{$\bm{u}^k \stackrel{i.i.d.}{\sim} h_i\left(\bm{u}\right)$}
                \While{\textit{true}}
                \Comment SIS loop
                \State $i \gets i+1$
                \State Sample $\mathbf{U}_{\mathrm{tmp}} = \{\bm{u}^k\}_{k=1}^{n_{\mathcal{E}}} \in \mathbb{R}^{n_{\mathcal{E}} \times d}$
                \Comment{$\bm{u}^k \stackrel{i.i.d.}{\sim} h_i\left(\bm{u}\right)$}
                \State Compute $\mathbf{G}_{\mathrm{tmp}} = G(\mathbf{U}_{\mathrm{tmp}}) \in \mathbb{R}^{n_{\mathcal{E}} \times 1}$
                \State Append $\mathbf{U}_{\mathcal{E}} \gets [\mathbf{U}_{\mathcal{E}},\mathbf{U}_{\mathrm{tmp}}]$
                \State Append $\mathbf{G}_{\mathcal{E}} \gets [\mathbf{G}_{\mathcal{E}},\mathbf{G}_{\mathrm{tmp}}]$
                \If{$i > 1$}
                \State Run $\widehat{G} = $             PLS-PCE$(\mathbf{U}_{\mathcal{E}},\mathbf{G}_{\mathcal{E}},p,\mathrm{'W'})$
                \Comment \cref{alg:PLS-PCE}
                \State Run $\mathbf{U}_{i-1},\bm{\mathrm{G}}_{i-1} = $
                SIS-aCS$(\widehat{G} ,\delta_{\mathrm{target}},n,d,i-1)$
                \Comment  \cref{alg:SIS-aCS}
                \EndIf
                \State Run $\mathbf{U}_{\mathcal{E}},\mathbf{G}_{\mathcal{E}}, \widehat{G} = $ Active Learning($G\left(\bm{u}\right)$, $\epsilon_{\mathrm{AL}}$, $n_{\mathrm{add}}$, $p$, $\mathbf{U}_{\mathcal{E}}$,$\mathbf{G}_{\mathcal{E}}$, $ \mathbf{U}_{i-1}$)
                \Comment \cref{alg:AL}
                \State Compute $\mathbf{G}_{i-1} = \widehat{G}(\mathbf{U}_{i-1}) \in \mathbb{R}^{n \times 1}$
                \State Compute $\sigma_i$ according to \cref{eq:sigma_i}
                \State Compute $\widehat{\omega}_i$ and $\widehat{s}_i$ according to \cref{eq:pf_ssis_aux}
                \State $\mathbf{U}_{i-1},\mathbf{G}_{i-1}$ $\gets$ resample from $\mathbf{U}_{i-1},\mathbf{G}_{i-1}$  with weights $\widehat{\omega}_i(\mathbf{U}_{i-1})$
                \Comment sample with replacement
                \State Run $\mathbf{U}_{i},\bm{\mathrm{G}}_{i} = $ SIS-aCS$(\mathbf{U}_{i-1},\bm{\mathrm{G}}_{i-1})$
                \Comment Perform a single MCMC step
                \If{\cref{eq:exit_crit_ssis_al}}
                \State Set $M \gets i$
                \State Run $\mathbf{U}_{\mathcal{E}},\mathbf{G}_{\mathcal{E}}, \widehat{G} = $ Active Learning($G\left(\bm{u}\right)$, $\epsilon_{\mathrm{AL}}$, $n_{\mathrm{add}}$, $p$, $\mathbf{U}_{\mathcal{E}}$,$\mathbf{G}_{\mathcal{E}}$, $ \mathbf{U}_{i-1}$)
                \Comment \cref{alg:AL}
                \State \textbf{break}
                \EndIf
                \EndWhile
                \State Run $(\mathbf{U}_{M},\bm{\mathrm{G}}_{M},\widehat{p}_{\mathrm{ASSIS}}) = $
                SIS-aCS$(\widehat{G}_{M} ,\delta_{\mathrm{target}},n,d,M)$
                \Comment  \cref{alg:SIS-aCS}
                \State \Return{$M,\mathbf{U}_{M},\bm{\mathrm{G}}_{M},\widehat{p}_{\mathrm{ASSIS}}$}
        \end{algorithmic}
\end{algorithm}

%%%%%%%%%%%%%%%%%%%%%%%%%%%%%%%%%%%%%%%%%%%%%
\section{Numerical experiments}
\label{sec:numerical_experiments}
\subsection{Error measures}
\label{sec:error_measures}
In the following, we examine a series of examples of low to high input dimensionality characterized by varying degrees of nonlinearity of the LSF and varying number of disconnected failure regions. The computational cost of each approach is measured with the total number of required calls to the underlying computational model.
The accuracy of the estimator is measured in terms of relative bias and CoV:
\begin{align}
\label{eq:bias}
\mathrm{relative~bias} &= \frac{p-\mathbb{E}[\widehat{p}]}{p},\\
\label{eq:cov}
\mathrm{CoV} &= \frac{\sqrt{\mathbb{V}[\widehat{p}]}}{\mathbb{E}[\widehat{p}]},
\end{align}
where $p$ is the known exact probability of failure or a reference solution computed with a large number of samples as reported in the corresponding references in \cref{tab:example_low_d}.
Further, we compute the relative root mean squared error (RMSE) of the probability of any failure estimate $\widehat{p}$, which combines bias and variability of the estimator as
\begin{equation}
\label{eq:rmse}
\mathrm{relative~RMSE} = \sqrt{\frac{\mathbb{E}[(p-\widehat{p})^2]}{p^2}} = \sqrt{\mathrm{relative~bias}^2 + \left(\frac{\mathbb{E}[\widehat{p}]}{p}\right)^2 \mathrm{CoV}^2}.
\end{equation}
The expectation and variance operators in the above equations are approximated by repeating each analysis $100$ times.
Additionally, the relative estimation error is defined as
\begin{equation}
\label{eq:rel_err}
\mathrm{relative~error} = \frac{\widehat{p}}{p}.
\end{equation}\pagebreak{}

%%%%%%%%%%%%%%%%%%%%%%%%%%%%%%%%%%%%%%%%%%%%%
\subsection{Low- and medium-dimensional examples}
The subspace importance sampler is designed to tackle high-dimensional problems, yet its performance should not deteriorate as the problem dimension decreases. We first investigate its performance in eight exemplary problems with dimension $2 \le d\le 100$. We demonstrate how both SSIS and ASSIS cope with multiple failure domains, strong nonlinearities, and extremely small target failure probabilities. In the interest of brevity, the examples are listed in \cref{tab:example_low_d} along with the problem dimension, target probability of failure, and key characteristics of the problem. The references provided in \cref{tab:example_low_d} may be consulted for detailed descriptions of the problem setups.
% Table 1
\begin{table}
\setlength{\tabcolsep}{.25pc}
        \centering\footnotesize
        \caption{Low- to medium-dimensional investigated benchmark problems.}
        \label{tab:example_low_d}
        \scriptsize{\tabcolsep2pt
                \begin{tabular}{lccccc}
                        \hline
                        Problem & Failure probability &  Inputs &  Input variables& Properties & References \\
                        \hline
                        \hline
                        Hat & $1.037\cdot 10^{-4} $ & $2$  & Standard-normal & Strongly nonlinear  & \cite{Schoebi2017}\\
                        \hline
                        Cantilever & $3.94\cdot 10^{-6} $ & $2$  & Gaussian &  Strongly nonlinear & \cite{Bect2017}\\
                        \hline
                        4-Branch & $5.60\cdot 10^{-9}$ & $2$  & Standard-normal &  Multiple failure regions; & \cite{Bect2017,Waarts2000}    \\
                        (acc. to \cite{Bect2017})&&&&extremely rare event &\\
                        \hline
                        Borehole & $ 1 \cdot 10^{-5}$ & $8$  & Log-normal, &  Strongly nonlinear, no underlying& \cite{An2001}    \\
                        ($276.7 \frac{m^3}{\mathrm{year}}$) &&&uniform&low-dimensional structure&\\
                        \hline
                        %               Oscillator & $3.78\cdot 10^{-7}$ & $8$  & Log-normal &  Strongly nonlinear;& \cite{DeStefano1991} \\
                        %               &&&&difficult for PCEs &\\
                        Truss & $1.6 \cdot 10^{-3}$ & $10$  & Log-normal, & Mildly nonlinear &\cite{Lee2006} \\
                        ($0.12$m) &&&Gumbel&& \\
                        \hline
                        Rare truss & $1.02\cdot 10^{-8}$ & $10$  & Log-normal, & Extremely rare event;& \cite{Lee2006} \\
                        ($0.18$m) &&&Gumbel& nonlinear & (modified)\\
                        \hline
                        Quadratic & $6.62\cdot 10^{-6}$ & $10$  & Standard-normal & Strongly nonlinear; underlying & \cite{Engelund1993,Uribe2020}\\
                        ($\kappa = 5$)&&&&low-dimensional structure&\\
                        \hline
                        Quadratic & $6.62\cdot 10^{-6}$ & $100$  & Standard-normal & Strongly nonlinear; underlying & \cite{Engelund1993,Uribe2020}\\
                        ($\kappa = 5$)&&&&low-dimensional structure&\\
                        \hline
                \end{tabular}
        }
\end{table}

\begin{figure}[h!]
        \centering
        \includegraphics[width=\textwidth]{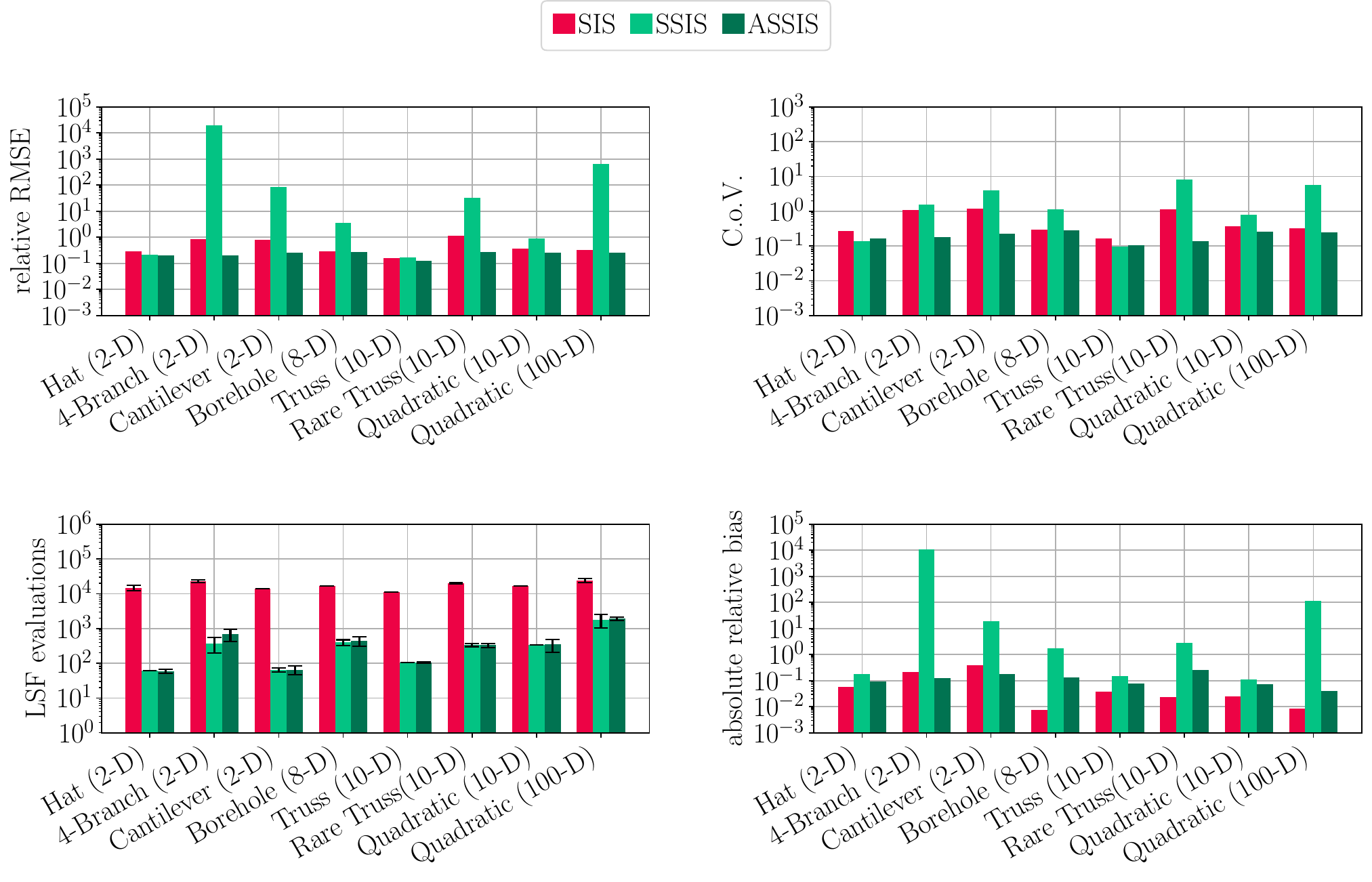}
        \caption{Low- and medium-dimensional examples: accuracy and cost comparison. Cost error bars include $\pm$ $2$ standard deviations.}
        \label{fig:performance}
\end{figure}

We solve the example problems with SIS-aCS with $n = 2\cdot 10^3$ samples per level and a burn-in period of $b=5$ samples within each MCMC chain.
As suggested in \cite{Papaioannou2016}, we choose $\delta_{\mathrm{target}} = 1.5$ for the exit criterion \cref{eq:exit_crit} for SIS-aCS as well as our surrogate-based samplers.
We compare this reference to SSIS and ASSIS for which we use an initial sample size of $n_{\mathcal{E}} = 5d$. All underlying PLS-PCE-W models are computed with a maximum number of subspace directions of $m=10$ and a maximum total polynomial degree of $|q|_{\ell_{q}} \le 7$, where $q=0.75$. To achieve a fair comparison between ASSIS and SSIS, we run first ASSIS and then SSIS with $n_{\mathcal{E}}$ for the latter chosen such that both methods use an approximately equal number of LSF evaluations. For both SSIS and ASSIS, we choose $n = 10^4$ with a burn-in period of $b = 30$. For ASSIS, we set $\epsilon_{\mathrm{AL}} =0.1$. Within SSIS/ASSIS many samples per level and long burn-in periods are affordable as sampling is performed with the surrogate model. For ASSIS we select $n_{\mathrm{add}} = 1$ unless prior knowledge of the problem structure suggests otherwise (the only exception in the set of examples considered here is the 4-branch function for which we select $n_{\mathrm{add}} = 4$ as it features four relevant failure regions in the input space). Figure~\ref{fig:performance} displays the performance of SIS, SSIS, and ASSIS for the examples in \cref{tab:example_low_d} in terms of the error measures defined in \cref{eq:bias,eq:cov,eq:rmse} and the total number of LSF evaluations (with the original model).\enlargethispage*{1pc}
% Figure 4

For all showcased examples, ASSIS yields equally or more accurate estimates compared to SSIS at equal cost. It also either matches or outperforms SIS at significantly reduced costs. Except for the easiest problems, i.e., those featuring well-behaved (truss) or low-dimensional (two-dimensional hat) LSFs associated with comparatively large failure probabilities, the in-level adaptivity of ASSIS leads to significant bias correction (Figure~\ref{fig:performance}, bottom right) and variance reduction (Figure~\ref{fig:performance}, top right).

\cite{Papaioannou2016} discusses the choice of the MCMC sampler for SIS and Finds that aCS as employed here is outperformed by a Gaussian mixture proposal in low-dimensional problems, while the latter is the preferred choice as the problem dimension grows. Our method is designed for the solution of high-dimensional reliability problems, and we thus consistently use aCS.
%%%\deleted[id=2]{However, we remark that in the 2-dimensional examples the relative RMSE produced by all considered SIS-based methods can be decreased by an order of magnitude by using the Gaussian %%%mixture proposal at constant cost.}

\begin{figure}[h!]
        \centering
        \includegraphics[width=\textwidth]{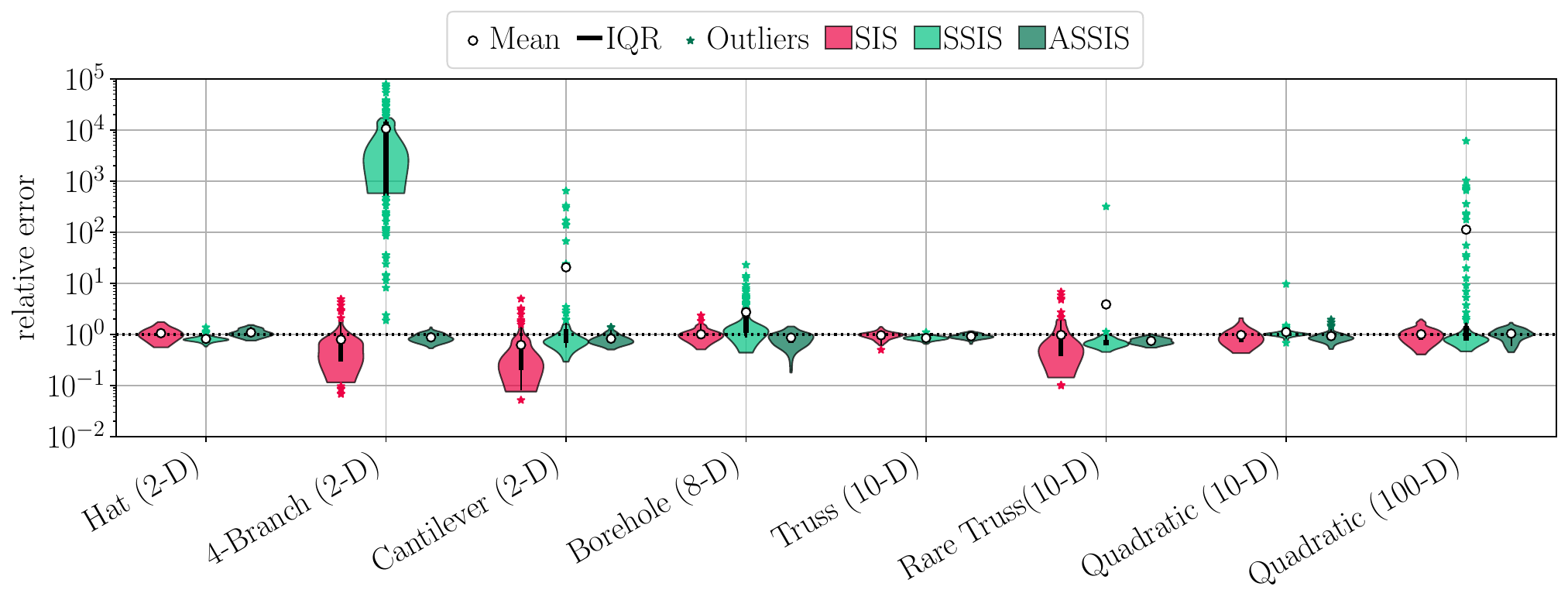}
        \caption{Low- and medium-dimensional examples: violin plots of the relative error along with means, interquartile ranges (IQR), and outliers. For the sake of clarity, kernel density estimates are computed after excluding outliers based on the relative distance to the data median.}
        \label{fig:rel_error}
\end{figure}

Comparing the truss and the rare truss models, the additional number of SIS levels required in the solution of the latter evidently leads to a deterioration of the SSIS estimate (Figure~\ref{fig:performance}, top left). This is due to single runs (less than 10 \%) among the 100 repetitions in which the sampled training sets lead to extreme outliers in the failure probability estimates (Figure~\ref{fig:rel_error}).
While this effect vanishes when increasing the number of samples in the training set, ASSIS offers a more cost-effective alternative to avoid such outliers by actively learning an informative augmentation of adverse training sets.
In this way, subspace identification and surrogate modeling errors cannot propagate and accumulate across the levels of SIS, as they are controlled by the AL procedure.
In fact, the phenomenon of rather rare but all the more severe outliers deteriorating the error mean and variability is a problem SSIS is facing not only in the rare truss example but also in the cantilever and both quadratic examples.
Conversely, it is seen that in the 4-branch example, SSIS consistently and considerably overestimates the probability of failure while ASSIS captures the probability of failure rather well.\enlargethispage*{1pc}
%%%\deleted[id=3]{The ASSIS-based probability of failure estimates each strive towards one of two accumulation points, one of which corresponds to the true probability of failure while the other %%%represents a slightly underestimated failure probability. This second cumulation point corresponds to ASSIS runs where only two of the four important failure regions of the 4-branch functions have been %%%identified.}\deleted[id=2]{This behaviour can be alleviated by employing a Gaussian mixture proposal in this example.}

% Figure 5
The two quadratic LSF models with 10 and 100 input dimensions demonstrate how the required number of LSF evaluations depends on the problem dimension in both surrogate-based approaches. This is due to the fact that the PLS-PCE model requires at least $d$ (often more) samples to identify a suitable subspace. Thus, as described above, we choose $n_{\mathcal{E}}$ as a multiple of $d$.
%%%\added[id=1]
{Since the surrogate-free version of SIS-aCS does not possess such a dependence on a problem dimension at all, the ratio of computational cost associated with SIS and ASSIS decreases as $d$ increases. This observation also indicates that if $d$ grows large enough, SIS-aCS will outperform any surrogate-based approach. This is expected for cases with $d = \mathcal{O}(10^5)$ and above; therefore, this observation is of little practical relevance for most engineering models, where ASSIS will likely be the most cost-effective choice.}

\enlargethispage*{1pc}
%%%%%%%%%%%%%%%%%%%%%%%%%%%%%%%%%%%%%%%%%%%%%
\subsection{High-dimensional example: Steel plate}
\begin{figure}[!t]
        \centering
        \begin{minipage}[t]{0.5\textwidth}
                \raisebox{-.25cm}{\includegraphics[width=\textwidth]{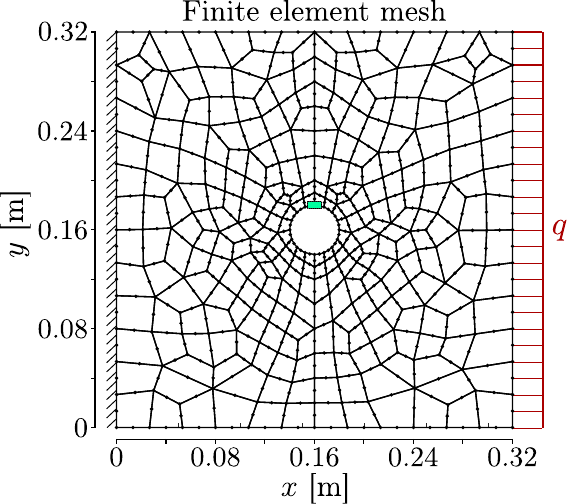}}
        \end{minipage}
        \hspace{0.5cm}
        \begin{minipage}[b]{0.42\textwidth}
                \subfloat{\includegraphics[width=0.33\textwidth]{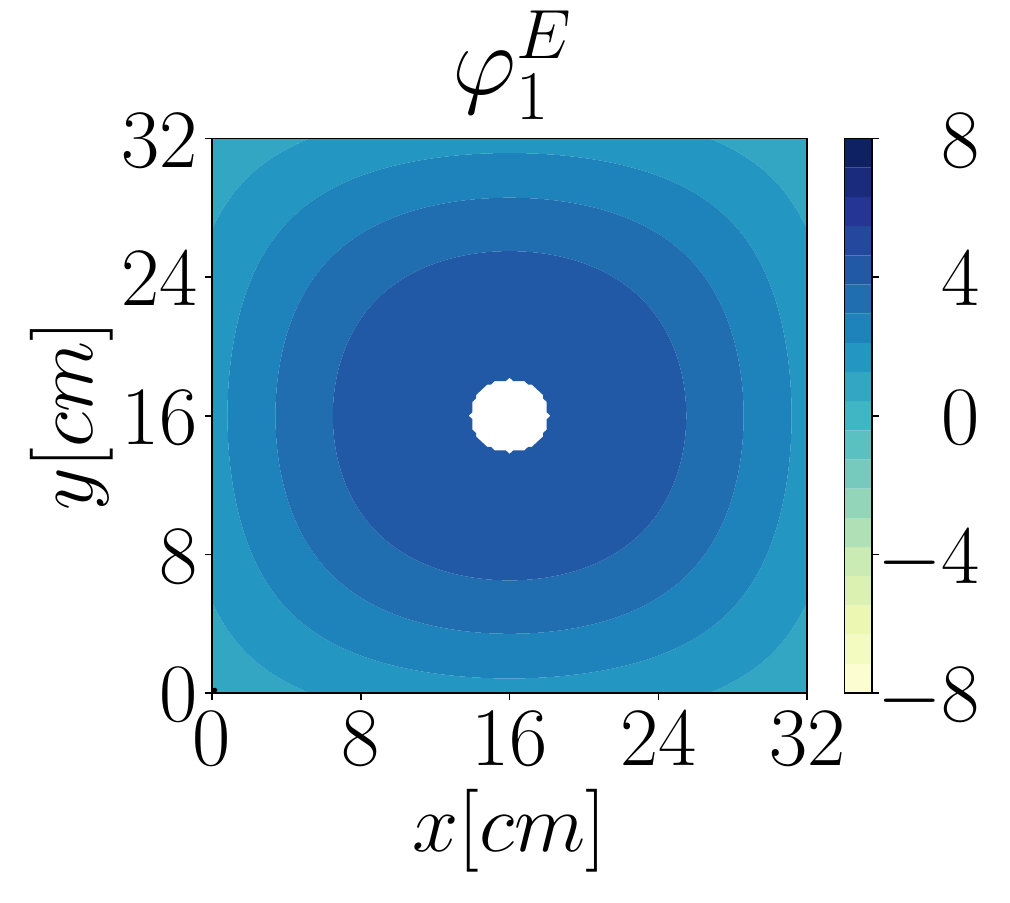}}
                \hfill
                \subfloat{\includegraphics[width=0.33\textwidth]{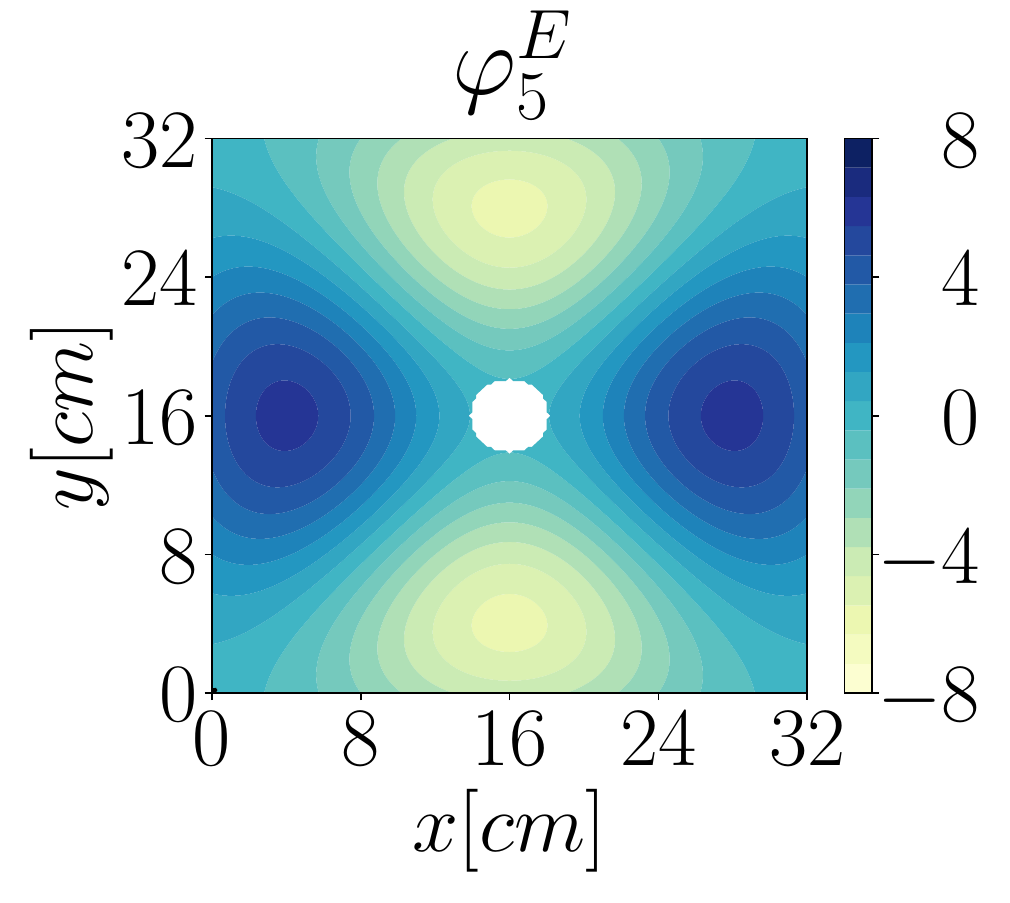}}
                \hfill
                \subfloat{\includegraphics[width=0.33\textwidth]{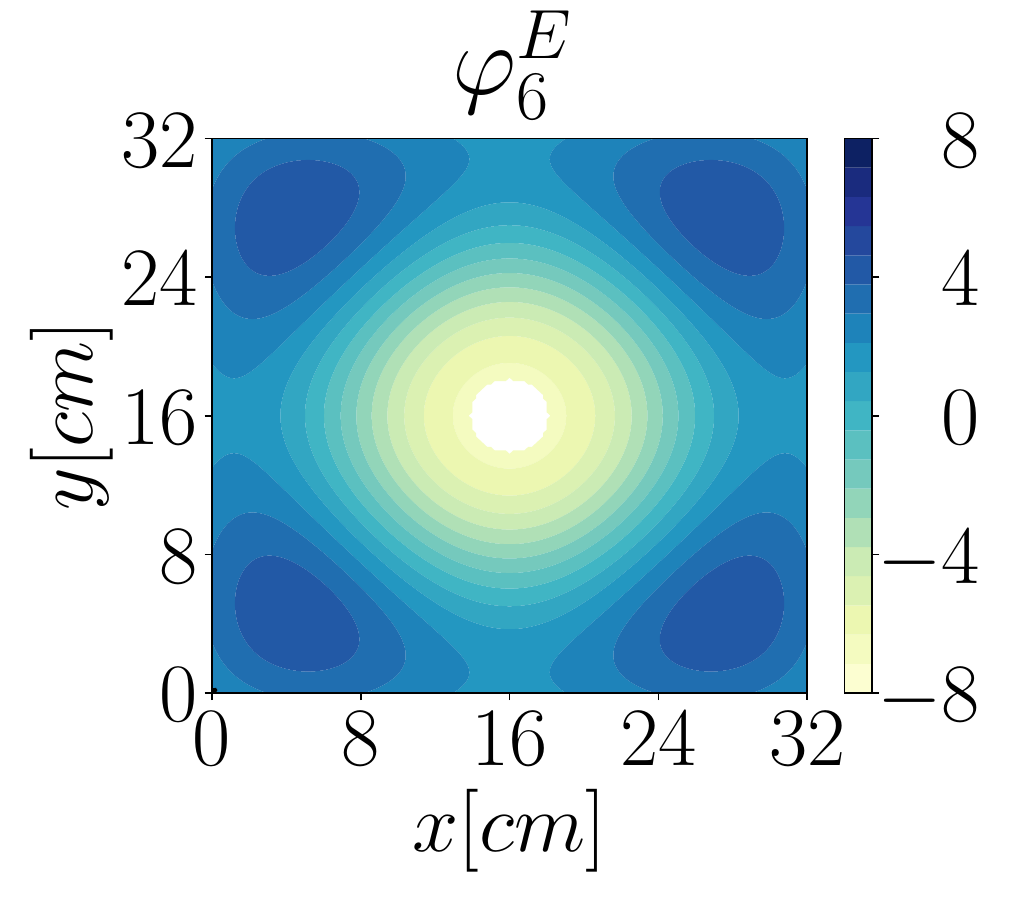}}
                \\
                \subfloat{\includegraphics[width=0.33\textwidth]{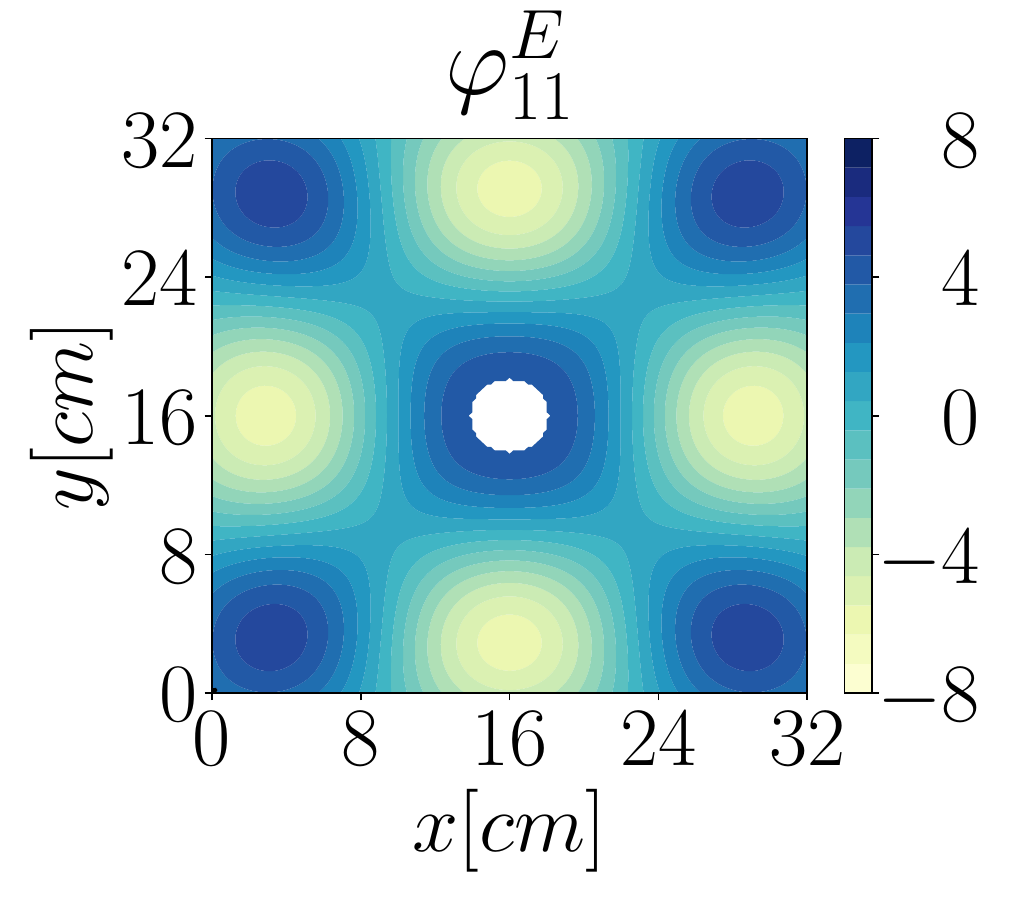}}
                \hfill
                \subfloat{\includegraphics[width=0.33\textwidth]{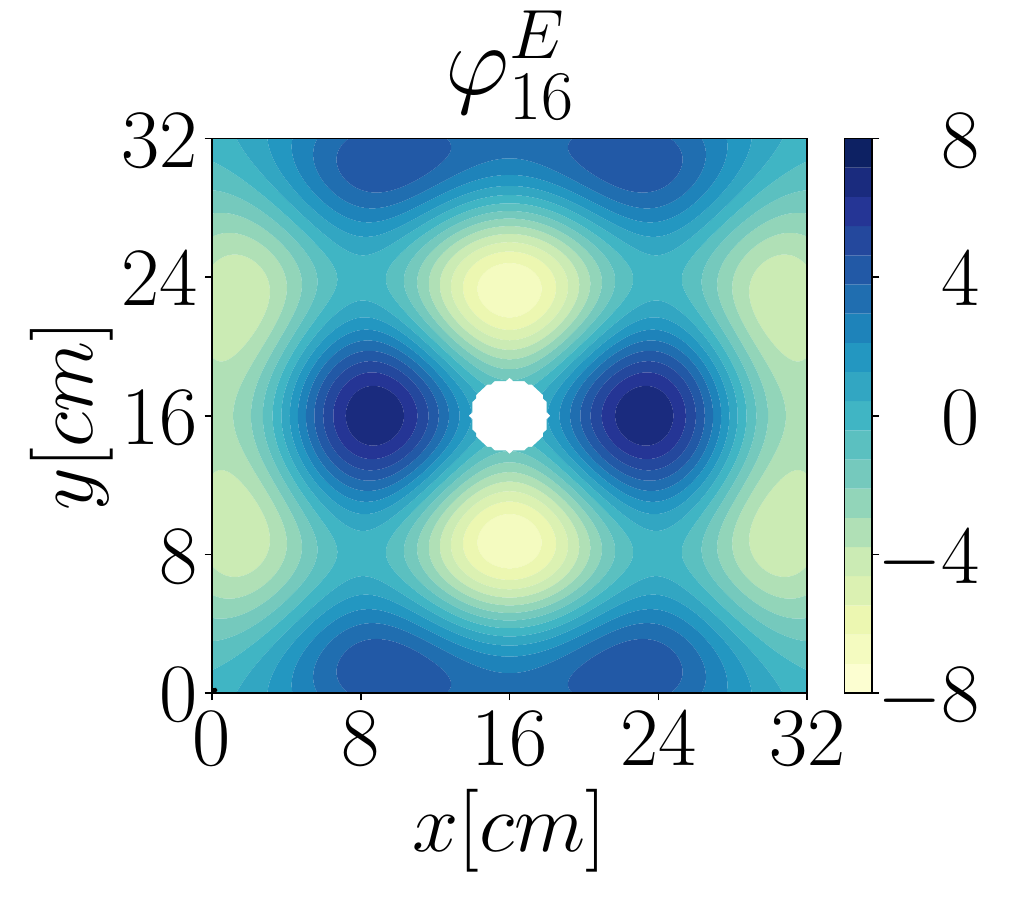}}
                \hfill
                \subfloat{\includegraphics[width=0.33\textwidth]{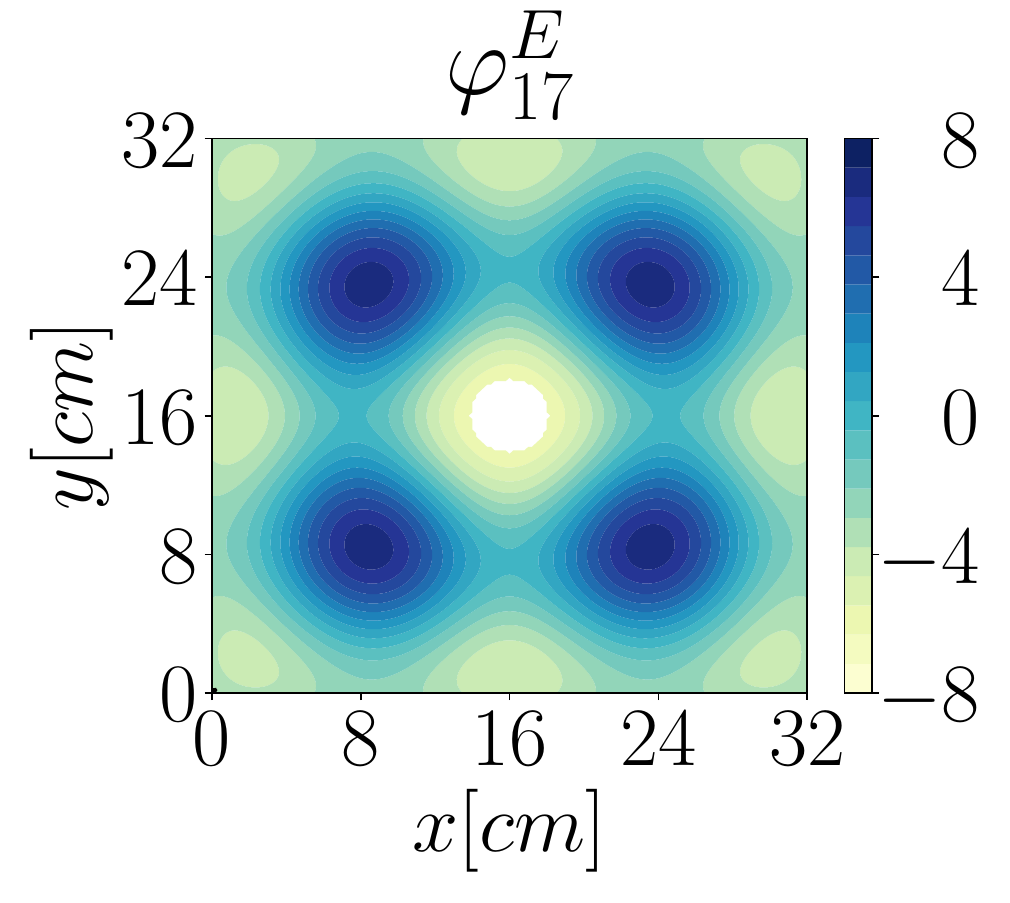}}
                \\
                \subfloat{\includegraphics[width=0.33\textwidth]{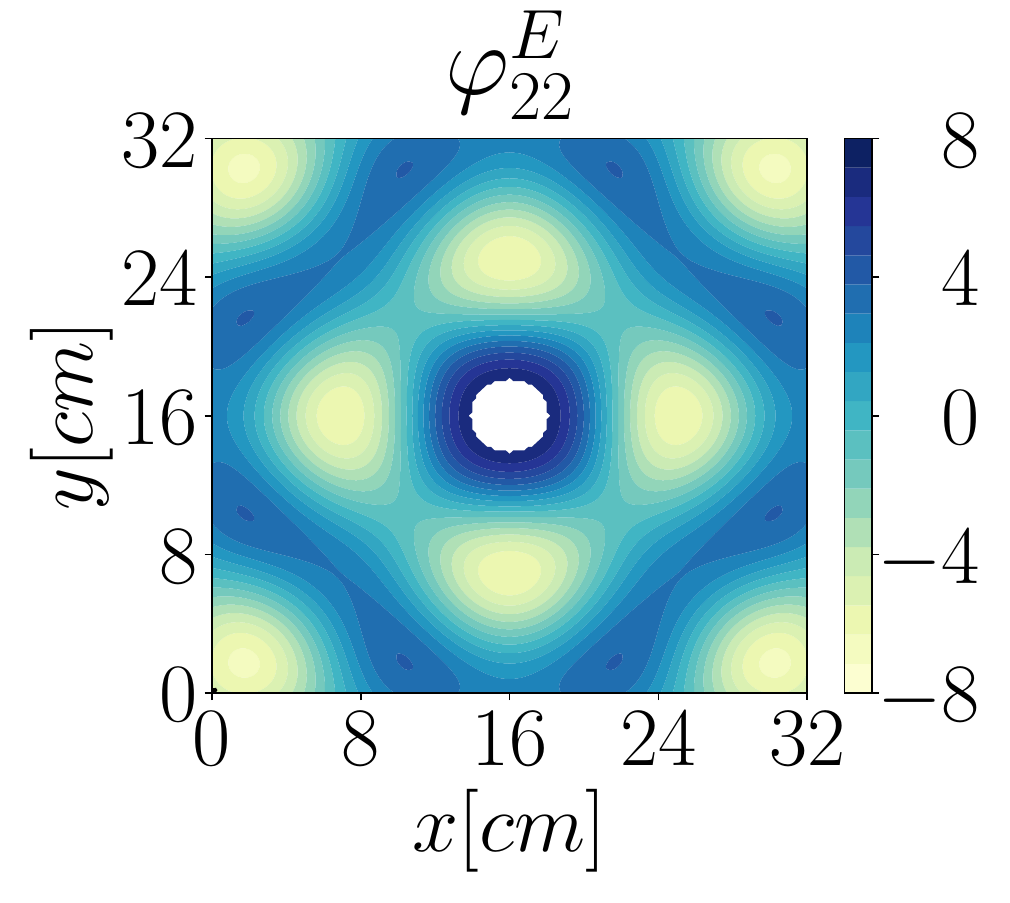}}
                \hfill
                \subfloat{\includegraphics[width=0.33\textwidth]{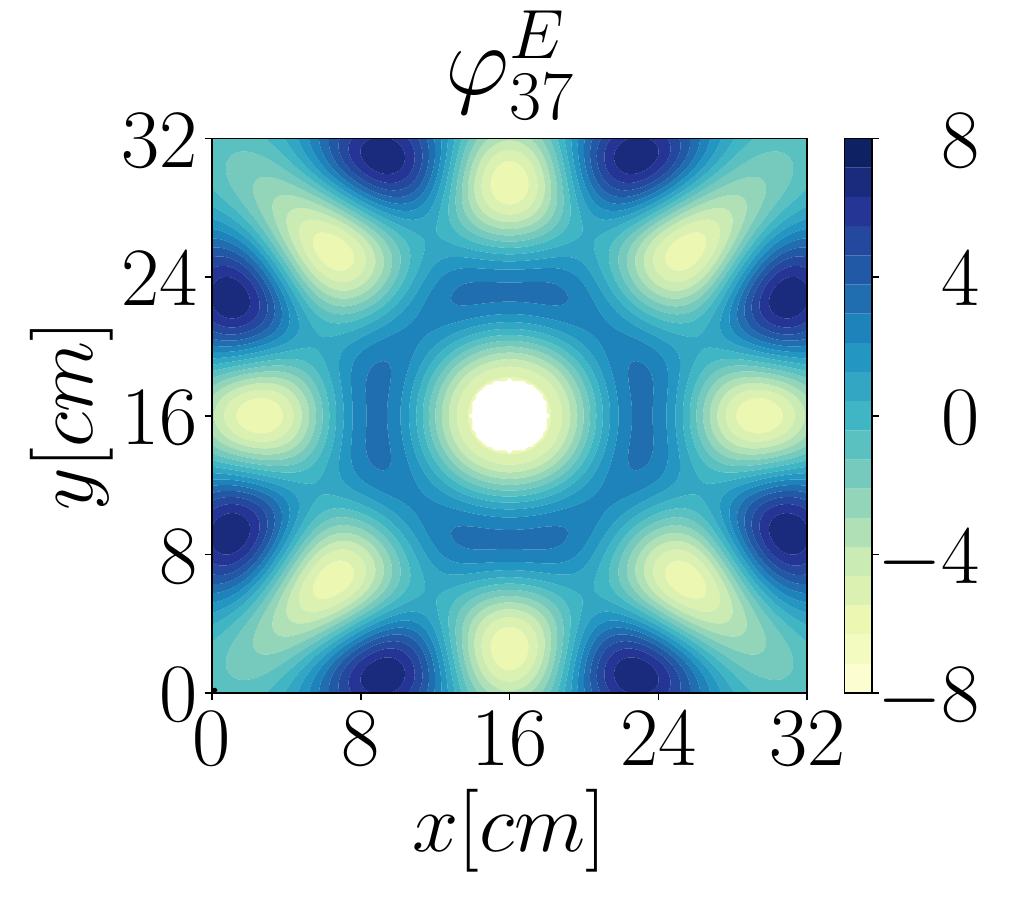}}
                \hfill
                \subfloat{\includegraphics[width=0.33\textwidth]{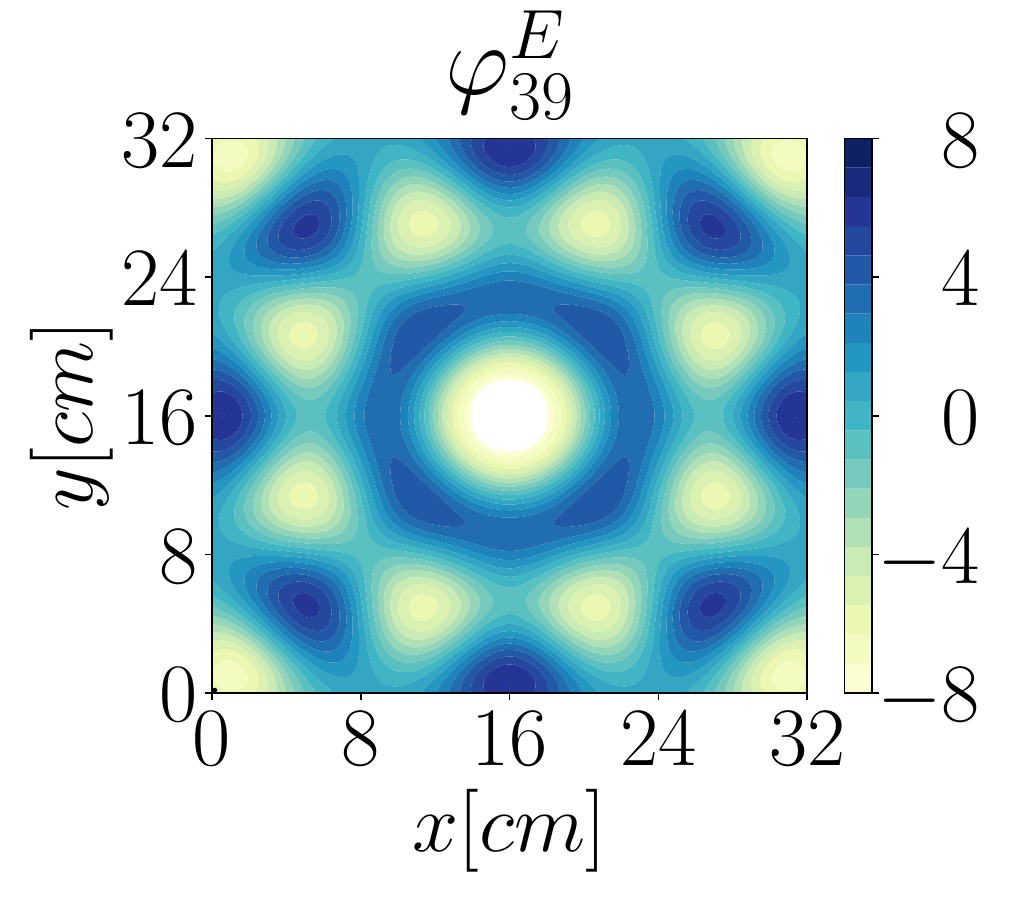}}
        \end{minipage}
        \caption{Left: Finite element mesh of  two-dimensional plate model with control node of the first principal stress $\sigma_1$.}
        \label{fig:plate}
\end{figure}

We consider a modified version of the example given in \cite{Uribe2020,liu_and_liu_1993}, which consists of a low-carbon steel plate of length $0.32$~m, width $0.32$~m, thickness $t = 0.01$~m, and a hole of radius $0.02$~m located at the center. The Poisson ratio is set to $\nu = 0.29$, and the density of the plate is $\rho=7850$~kg/m$^3$. The horizontal and vertical displacements are constrained at the left edge. The plate is subjected to a random surface load that acts on the right narrow plate side. The load is modeled as a log-normal random variable with mean $\mu_q = 60$ MPa and $\sigma_q = 12$ MPa. The Young's modulus $E(x,y)$ is considered uncertain and spatially variable. It is described by a homogeneous random field with lognormal marginal distribution, mean value $\mu_E = 2\times 10^{5}$~MPa, and standard deviation $\sigma_E = 3\times 10^{4}$~MPa.
The autocorrelation function of the underlying Gaussian field $\ln E$ is modeled by the isotropic exponential model
\begin{equation}
\label{eq:corr_E}
\rho_{\ln E}(\Delta x,\Delta y) = \exp \left\{-\frac{\sqrt{\Delta x^2 + \Delta y^2}}{l_E}\right\}
\end{equation}
with correlation length $l_{\ln E} = 0.04 \text{m}$.
The Gaussian random field $\ln E$ is discretized by a Karhunen--Lo\`eve expansion with $d_E=868$, which yields a mean error variance of 7.5\% and reads
\begin{equation}
\label{eq:KL_E}
E(x,y) = \exp\left\{\mu_{\ln E} + \sigma_{\ln E} \sum_{i=1}^{d_E} \sqrt{\lambda^E_i} \varphi^E_i(x,y) \xi_i\right\}.
\end{equation}
$\mu_{\ln E}$ and $\sigma_{\ln E}$ are the parameters of the log-normal marginal distribution of $E$, $\{\lambda^q_i,\varphi^E_i\}$ are the eigenpairs of the correlation kernel in $\cref{eq:corr_E}$, and $\bm{\xi} \in \mathbb{R}^{d\times1}$ is a standard-normal random vector.
The most influential eigenfunctions (based on a global output-oriented sensitivity analysis of the plate model performed in \cite{Ehre2020b}) are shown in Figure~\ref{fig:plate} on the right.

% Figure 6

The stress ($\bm{\sigma}(x,y) = [\sigma_x(x,y),\sigma_y(x,y),\tau_{xy}(x,y)]^{T}$), strain ($\bm{\epsilon}(x,y)=[\epsilon_x(x,y),$ $\epsilon_y(x,y),\gamma_{xy}(x,y)]^{T}$), and displacement ($\mathbf{u}(x,y)=[u_x(x,y),u_y(x,y)]^{T}$) fields of the plate are given through elasticity theory, namely, the Cauchy--Navier equations \cite{johnson_2009}. Given the configuration of the plate, the model can be simplified under the plane stress hypothesis, which yields
\begin{equation}\label{eq:plate}
G(x,y)\nabla^2\mathbf{u}(x,y) + \frac{E(x,y)}{2(1-\nu)} \nabla(\nabla\cdot\mathbf{u}(x,y)) + \mathbf{b} =0.
\end{equation}
Therein, $G(x,y) := E(x,y)/(2(1+\nu))$ is the shear modulus, and $\mathbf{b} = [b_x,b_y]^{T}$ is the vector of body forces acting on the plate. \cref{eq:plate} is discretized with a finite-element method. That is, the spatial domain of the plate is discretized into $282$ eight-noded quadrilateral elements, as shown in Figure~\ref{fig:plate}.
%%%\added[id=2]
{In a grid independence study, the plate's probability of failure was found to slightly increase with decreasing mesh element size, which is likely due to the reduction of averaging effects when integrating higher-order Karhunen--Lo\`eve terms. However, for the purpose of testing ASSIS, the model is sufficiently accurate and features two important properties: 1. It possesses a low-dimensional structure that can be exploited with dimensionality-reducing surrogates. 2. It is truly high-dimensional in the sense that the solution does not only depend on a small subset of the input variables (i.e., the low-dimensional structure is not a trivial subspace of the original input space).}
The LSF is defined by means of a threshold for the the first principal plane stress
$$\sigma_1 = 0.5(\sigma_x + \sigma_y) + \sqrt{[0.5(\sigma_x + \sigma_y)]^2 + \tau_{xy}^2}$$ evaluated
at node 11 (see green marker in Figure~\ref{fig:plate}, left).
Node 11 indicates a location where maximum plane stresses occur frequently in this example.
The LSF reads
\begin{equation}
g(\bm{U}) = \sigma_{\mathrm{threshold}} - \sigma_1(\bm{U}),
\end{equation}
where $\sigma_{\mathrm{threshold}} = 450$~MPa. The target probability of failure is determined to $p = 4.23 \cdot 10^{-6}$ with $\mathrm{CoV} = 0.0119$ as the average of 100 repeated runs of subset simulation \cite{Au2001} with $10^4$ samples per level.

SIS-aCS is run with $n = 2\cdot 10^3$ samples per level and a burn-in period of $b=5$ samples within each MCMC chain.
SSIS and ASSIS are run with $n = 10^5$ samples per SIS level, a burn-in period $b=30$, and an AL threshold of $\epsilon_{\mathrm{AL}} =0.1$. In the first level $n_{\mathcal{E}} = 900$, and in each additional level only $n_{\mathcal{E}} = 100$ samples are added in the initial sampling phase.
\cref{tab:plate_results} lists the average estimated probabilities of failure along with error measures and average number of required LSF evaluations.
It is seen that both SSIS and ASSIS alleviate computational cost by more than an order of magnitude while at the same time reducing the relative RMSE by at least an order of magnitude. The decomposition of the RMSE in CoV and relative bias reveals that this is mostly due to variance reduction as SIS-aCS already yields a small bias.

% Table 2
\begin{table}[]
        \centering
        \caption{Accuracy and cost of SIS, SSIS \& ASSIS for the plate example based on $100$ repetitions of the analysis. The reference $ p_{\mathrm{ref}} = 4.23 \cdot 10^{-6}$ is computed with $100$ repeated runs of subset simulation with $10^4$ samples per level with $\mathrm{CoV} = 0.0119$ for the mean estimate.}
        \label{tab:plate_results}
        \footnotesize{
                \begin{tabular}{lccccc}
                        \hline
                        Method & $\mathbb{E}[p] $ &  relative RMSE & CoV & relative bias & avg. \# LSF evaluations\\
                        \hline
                        SIS-aCS & $3.88 \cdot 10^{-6}$ & $0.576$ & $0.625$ & $0.083$ & $17000$\\
                        SSIS & $3.99 \cdot 10^{-6}$ & $0.061$ & $0.021$ & $0.058$ & $1300$  \\
                        ASSIS & $4.10 \cdot 10^{-6}$  & $0.036$ & $0.021$ & $0.030$ & $1318$ \\
                        \hline
                \end{tabular}
        }
\end{table}
% Figure 7
\begin{figure}[t!]
        \centering
        \includegraphics[width=\textwidth]{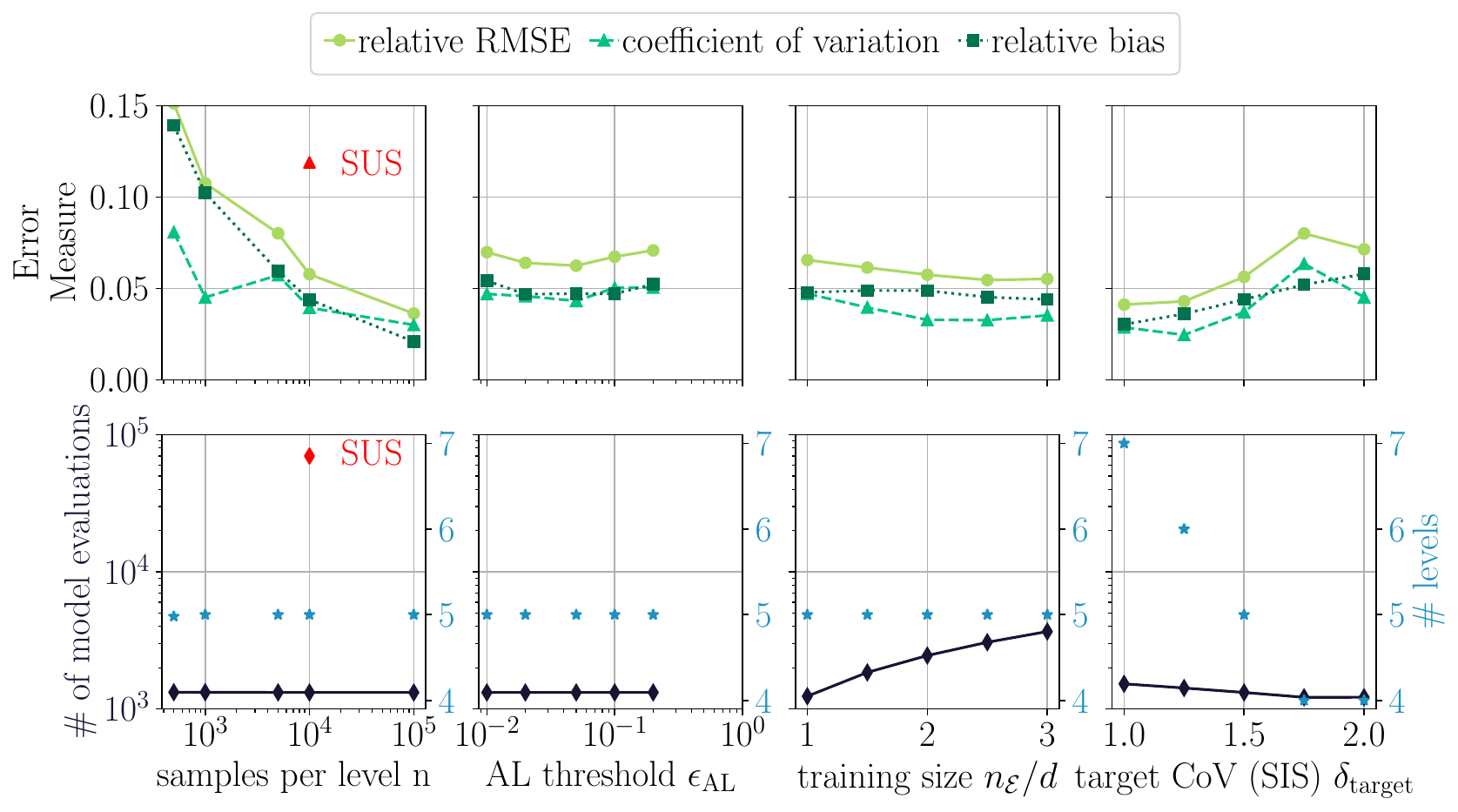}
        \caption{Steel plate reliability using ASSIS: parameter influence studies.
        %%%\added[id=1]
        {Top: Error measures as defined in \cref{eq:bias,eq:cov,eq:rmse} for ASSIS (green lines with markers). Bottom: Computational cost in terms of total number of LSF evaluations with the true computational model (left y-axis; black solid lines with diamond markers) and number of SIS levels to convergence (right y-axis; blue star markers). Top left: CoV of a subset simulation reference run with $n=10^{4}$ samples per level (red triangle marker). Bottom left: Total number of required LSF evaluations of a subset simulation (SUS) reference run with $n=10^{4}$ samples per level (red triangle marker).}}
        \label{fig:plate_parameter_study}
\end{figure}
A parameter study of important ``tweakable'' parameters of ASSIS is depicted in Figure~\ref{fig:plate_parameter_study}. Parameters that are not subject to a parametric study are chosen as above, with the exception of $n = 10^4$ instead of $n = 10^5$. The estimation error and computational cost of ASSIS are analyzed for varying AL threshold $\epsilon_{\mathrm{AL}}$, number of samples in the training set $n_{\mathcal{E}}$, the number of samples per SIS level $n$, and the target CoV $\delta_{\mathrm{target}}$ used for the SIS procedure. The scaling of 10\% between the initial training set and all subsequent training samples is kept constant.

The parameters  $\epsilon_{\mathrm{AL}}$ and $n_{\mathcal{E}}$ describe the behavior of the surrogate modeling and AL procedures, while $n$ and $\delta_{\mathrm{target}}$ describe SIS itself. Figure~\ref{fig:plate_parameter_study} shows that increasing the target CoV leads to a reduced number of levels in the SIS procedure, which is directly associated with a reduction in computational cost. The reduction is relatively small here as most of the samples are added in the first level.
By design, the number of required samples remains unaffected by varying the number of samples per SIS level, while the estimation error depends reciprocally on it. Conversely, and also by design, the computational cost depends monotonically on the choice of $n_{\mathcal{E}}$. If a majority of the used original LSF evaluations are added during an AL procedure, this relationship may be nonlinear. For the plate example, however, the initially drawn training samples at each level makes up for the majority of used original LSF evaluations, hence the linear dependency.
The estimation errors decrease slightly with increasing training set size, although the effect is limited as high accuracy is already achieved with the first training set of the lowest investigated %%%\deleted[id=2]{DoE}
size.The fact that the subspace does not change significantly with increasing SIS level leaves little to be learned by adding more LSF evaluations to the training set.
This is also the reason for the competitive performance of SSIS in this example.
The estimation errors (as well as the computational cost in this case) remain unaffected by varying AL thresholds $\epsilon_{\mathrm{AL}}$, which is in line with the observation that a large fraction of the computational budget is spent on
%%%\added[id=2]
{sampling} the initial training set rather than the AL-based training set augmentation.
%%%%%%%%%%%%%%%%%%%%%%%%%%%%%%%%%%%%%%%%%%%%%
\section{Concluding remarks}
\label{sec:conclusion}
This paper proposes a method for the cost-efficient solution of high-dimensional reliability problems.
We build on a recently introduced dimensionality-reducing surrogate modeling technique termed PLS-PCE \cite{Papaioannou2019} and previous work, in which we use PLS-PCE surrogates to reconstruct biasing densities within an  SIS \cite{Papaioannou2018b} (SSIS). We refine this approach by devising an AL procedure in each SIS level, which serves to effectively control the estimation error introduced by the surrogate-based importance density reconstructions. The learning procedure, i.e., the selection of new points for the training set, is driven by an estimate of both the subspace and surrogate model estimation error.
This criterion can be generally used in PCE-based AL procedures.

We showcase the performance of SSIS and ASSIS in nine example applications with input dimensionality ranging from $d=2$ to 869. The examples feature different typical caveats for reliability methods such as multiple failure domains, strongly nonlinear LSFs, and extremely small target probabilities of failure. Depending on the example, we achieve a cost reduction of one to over two orders of magnitude with ASSIS compared to the reference method (SIS with the original model) at equal or lower estimation errors. It is shown that SSIS is susceptible to the randomness of the initial training set  occasionally producing outliers if the training set is adverse. The AL procedure (ASSIS) remedies this drawback and stabilizes the estimator by augmenting potentially adverse training sets with informative additional samples.

The million dollar question, as with any surrogate model, is on the method's ability to generalize. Certainly, there exist examples that do not possess a suitable linear subspace as required by PLS-PCE modeling.
%%%\added[id=2]
{Further, cases of model misspecification may arise if the computational model cannot be represented with PCEs (e.g., if it is a rational function). Then, the probability of failure estimate produced by ASSIS will be neither consistent nor unbiased.}
However, by means of coupling PLS-PCE with SIS, both requirements are relaxed somewhat, as only a locally accurate surrogate model is required to propagate samples from one intermediate biasing density to the next. Hence, ASSIS can still be expected to perform well if the computational model may be represented in terms of a sequence of local linear subspaces on which the model can be approximated well with polynomials.
%%%\added[id=1]
{Relaxing the orthogonality or even the linearity assumption on the latent space transformation likely bears potential to improve the performance of dimensionality-reduced PCEs. Doing so will require methods to track the appropriate PCE basis upon determining the law of the transformed input random vector (as these will not be standard-normal if the latent space transformation is no longer subject to the orthogonality constraint).}
%%%%%%%%%%%%%%%%%%%%%%%%%%%%%%%%%%%%%%%%%%%%%
%%%\section{Acknowledgment}
%%%This project was supported by the German Research Foundation (DFG) through Grant STR 1140/6-1 under SPP 1886.

\enlargethispage*{1pc}

\end{document}